\documentclass{aastex}
\usepackage{emulateapj5}
\usepackage{natbib}

\shortauthors{Chiboucas, Barr, Flint, Jorgensen, Collobert, Davies}
\shorttitle{Gemini/HST Cluster project: the data}

\begin{document}

\title{The Gemini/HST Cluster Project: Structural and Photometric Properties of Galaxies in 
Three z $=0.28-0.89$ Clusters}

\author{Kristin Chiboucas\altaffilmark{1}}
\email{kchibouc@gemini.edu}

\author{Jordi Barr\altaffilmark{2}}
\email{jordi.barr@nottingham.ac.uk}
\author{Kathleen Flint\altaffilmark{3}}
\email{kflint@nationalpostdoc.org}
\author{Inger J{\o}rgensen\altaffilmark{1}}
\email{ijorgensen@gemini.edu}
\author{Maela Collobert\altaffilmark{2}}
\email{maela.collobert@gmail.com}
\author{Roger Davies\altaffilmark{2}}
\email{rld@astro.ox.ac.uk}

\altaffiltext{1}{Gemini Observatory, 670 N. A'ohoku Pl., Hilo, HI 96720}
\altaffiltext{2}{Department of Astrophysics, University of Oxford, Keble Rd, Oxford OX1 3RH, UK}
\altaffiltext{3}{National Postdoctoral Association, 1200 New York Avenue, NW, Suite 635, Washington, DC 20005}

\begin{abstract}
We present the data processing and analysis techniques we are using 
to determine structural and photometric properties of galaxies in our 
Gemini/HST Galaxy Cluster Project sample.  The goal of this study is
to understand cluster galaxy evolution in terms of scaling relations
and structural properties of cluster galaxies at 
redshifts $0.15 <$ z $< 1.0$.   To derive parameters such as total magnitude, 
half-light radius, effective surface brightness, and Sersic n,
we fit r$^{1/4}$ law and Sersic function 2-D surface brightness profiles to 
each of the galaxies in our sample.  Using simulated galaxies, we test how the assumed profile 
affects the derived parameters and how the uncertainties affect 
our Fundamental Plane results.  We find that while fitting galaxies
which have Sersic index n $< 4$ with r$^{1/4}$ law profiles systematically 
overestimates the galaxy radius and flux, the combination of 
profile parameters that enter the Fundamental Plane has uncertainties 
that are small. 
Average systematic offsets and associated random uncertainties 
in magnitude and $\log r_e$ for n $> 2$ galaxies 
fitted with r$^{1/4}$ law profiles
are $-0.1\pm0.3$ and $0.1\pm0.2$ respectively.
The combination of effective radius and surface brightness, 
$\log r_e - \beta \log \langle I \rangle_e$, 
that enters the Fundamental Plane produces
offsets smaller than $-0.02\pm0.10$. This systematic error is 
insignificant and independent of galaxy magnitude or size.  
A catalog of photometry 
and surface brightness profile parameters is presented for
three of the clusters in our sample, RX J0142.0+2131, RX J0152.7-1357, and 
RX J1226.9+3332 at redshifts 0.28, 0.83, and 0.89 respectively.

\end{abstract}


\section{Introduction}\label{intro}

Theoretical hierarchical models of galaxy formation predict that 
numerous small halos, primordial
dwarf galaxies, collapsed early on in the history of the universe.
These merged to form ever larger halos with deeper potentials
and greater baryonic mass accumulation.  While these $\Lambda$CDM models are
successful at explaining structure formation on large scales, the agreement 
with observations is worse on the level of galaxies \citep{con07}.
Understanding the formation of primordial galaxies and how 
galaxies came to have the stellar
populations, masses, and structural properties observed in
the local universe is fundamental to our understanding of cosmology.  

Generally, the physical processes a galaxy undergoes during its evolution
depend primarily on only two factors: mass and environment.  While mass
may affect galaxy structural and dynamical properties and star formation activity
in predictable ways, the local environment of the galaxy is the source of complex 
evolutionary processes such as
galaxy-galaxy merging, harassment, and stripping.
Despite this, tight scaling relations
of galaxy properties exist and must be reconciled with potentially different
evolutionary paths.  Such scaling relations include the Tully-Fisher 
relation between disk galaxy total magnitude and rotational velocity \citep{tf77}, 
the three-parameter Fundamental Plane (FP) relationship between early type 
galaxy velocity dispersion, effective radius, and effective surface brightness 
\citep[e.g.][]{dlbd87,dd87}, 
the red sequence for early type galaxies, and relationships
between absorption line strengths and central velocity dispersion.
How these tight relations evolve with redshift or vary with galaxy
mass places strong constraints on galaxy formation and
evolution models.

Observational studies of cluster galaxies have found that the most massive
galaxies are composed mainly of old stellar populations suggesting that massive
early type galaxies formed at redshifts $> 2$ \citep[e.g.][]{ble92,jorg99,tfwg00,blake03,mei06}.  
There is little evidence for recent star formation in these quiescent galaxies.
They form a red sequence with little scatter in a color-magnitude diagram \citep[e.g.][]{SV78}
and they obey tight scaling relations in their kinematic and structural properties
\citep[e.g.][]{dlbd87,jfk96}.   These all suggest an early epoch
of star formation in a homogeneously old galaxy population.  Although massive
galaxies exhibit pure passive evolution since z $\sim2$ evolving only in
luminosity and color as stellar populations age, observations have also shown
that clusters and lower mass cluster galaxies continue to evolve at intermediate
redshifts.  The fraction of blue, star forming galaxies increases with
redshift \citep[e.g.][]{BO84, Ellingson01}  while larger fractions of spiral to S0 galaxies are
observed in clusters at intermediate redshifts \citep[e.g.][]{Dressler97}.  

Meanwhile
hierarchical structure formation models predict that galaxies form from the collective 
mergers of smaller galaxies. 
These mergers are expected to continue with high frequency through 
intermediate redshifts \citep[e.g.][]{Baugh96}.  Semi-analytical
models predict a characteristic mass a factor of 3 lower 
than found at z = 1 \citep{poggianti04} with simulations underpredicting the 
numbers and mass densities of the
most massive galaxies by a factor of over 100 \citep{con07}, 
all implying that massive galaxies undergo a faster build-up than predicted.
Dynamical
models also often show a greater range in early type galaxy structure and kinematics 
than observed \citep{deZF91}.

In order to reconcile apparently contradictory observations with formation and evolution 
models, we must study galaxy populations over a wide range of redshifts and
for a wide range in galaxy mass.
Previous studies have tended to investigate large samples of galaxies within 
a single cluster or epoch \citep[e.g.][]{fz05, FPz0}, or fewer galaxies 
over a range of redshifts \citep[e.g.][]{tsc02}, 
and often only the brightest, most massive galaxies.  Through the
Gemini/HST Galaxy Cluster Project, we seek to better understand
the processes driving cluster galaxy evolution by studying scaling relations
as a function of mass and environment since z $= 1.0$, at a time when the universe 
was only half its current age.  We have therefore obtained data for a
large sample of galaxies, which
includes a sufficient number of galaxies at each redshift
to accurately measure scaling relations and a sufficient
number of redshifts to measure trends in the scaling
relations as a function of redshift \citep{jorg06,jorg07}.  At each redshift 
our sample spans a
wide range in galaxy luminosity in order to investigate the role of galaxy mass.
To discriminate between
evolutionary effects due to environment and due to galaxy mass we have obtained
consistent radial coverage for each cluster.

In a previous paper we have described the project and sample \citep{jorg05}.
Our sample of galaxy clusters consists of 15 
X-ray selected massive clusters with redshifts ranging from $0.15 < z < 1$.  
For each of the clusters in our sample, we have obtained Gemini/GMOS 
spectra for $\sim 30$ cluster members with a wide range of luminosities.  
The targetted objects are chosen independently of morphology since there is
evidence that morphologies may continue to evolve even to the present epoch 
and we wish to avoid ``progenitor bias" \citep{vdf01}.  
From the spectra, we obtain redshifts, velocity dispersions, and line index 
measurements.  Using HST imaging data, we measure structural parameters and surface 
brightness profiles.  
With this large sample, we are studying scaling relations such as line index strengths 
and the FP as
a function of redshift.   Line index measurements probe chemical enrichment
providing insight into the star formation histories of the galaxies.
The FP is examined in terms of mass and mass-to-light (M/L) ratios in order to probe both
assembly histories and luminosity evolution.   We are using structural parameters
for quantitative measurements of morphology to study morphological evolution.
We have published the results on line index relations for
RX J0152.7-1357 \citep{jorg05} and RX J0142.0+2131 \citep{barr05}, and the 
FP for RX J0142.0+2131 \citep{barr06} and RX J0152.7-1357 and 
RX J1226.9+3332 \citep{jorg06,jorg07}.

The FP relates surface brightness, effective radius, and velocity dispersion
in a tight and well established relation \citep{dlbd87,dd87,jfk96}  
\begin{equation}
\log r_{e} = \alpha \log\sigma + \beta \log\langle I\rangle_{e} + \gamma.
\end{equation}
It can also be described
as a relation between M/L and $\sigma$ or as
\begin{equation}
\log (M/L) = \xi \log M + \gamma^{\prime}
\end{equation}
which makes the FP a powerful tool to study the
star formation and assembly history in early type galaxies.
A comparison of the FP for our high redshift sample to that of low redshift 
Coma cluster galaxies
has revealed a number of interesting results.  The two z $\sim 0.85$ clusters 
exhibit a steeper slope than the low redshift FP. We find this to be evidence for
downsizing in which the lower mass galaxies have undergone more recent star
formation and are overluminous compared to their low-z counterparts \citep{jorg06,jorg07}.
The FP for RX J0142.0+2131 at z=0.28, while having the same slope as the low
z sample, displays a greater scatter, possible evidence for the galaxies having
undergone rapid bursts of star formation during a cluster merger 
at z $> 0.85$ \citep{barr06}.

Future papers will address the FP, line index
scaling relations, and galaxy quantitative morphologies for all clusters in our sample
in order to compare the star formation and assembly histories over a range
of redshifts.
Here we present the data reduction and analysis techniques we have used 
to measure structural and photometric parameters for 
three clusters in our sample, RX J0152.7-1357, RX J1226.9+3332, and RX J0142.0+2131.

RX J0152.7-1357 is a massive cluster at z=0.83 originally discovered from 
ROSAT data.  XMM-Newton and
Chandra observations later showed the cluster to consist of two subclumps in the
early stages of merging \citep{Jones04,Maughan03,Girardi05}.   Much research
has been done on this cluster, including recent studies of star formation rates
\citep{homeier05}, morphology, and the color-magnitude diagram \citep{blake}.
RX J1226.9+3332 is a massive cluster at z=0.89 and also the most X-ray luminous 
cluster known at such high redshift.  It was discovered  in the 
Wide Angle ROSAT Pointed Survey \citep{ebe01} and exhibits a relaxed
morphology. However,
in a deep XXM-Newton and Chandra study of the X-ray mass analysis of this
cluster, \cite{MJJV07} find evidence for a recent or ongoing merger event.
RX J0142.0+2131, at z=0.28, was first identified as a massive cluster
in both the Northern ROSAT All-Sky Galaxy Cluster Survey \citep{BVH00} and 
the ROSAT extended Brightest Cluster Sample \citep{Ebe00}.   Although
it is a relatively poor cluster, it displays
a large cluster velocity dispersion, yet shows no signs 
of substructure \citep{barr05}.

In this paper we describe our measurements of the physical properties of the galaxies
in our sample and analyze how uncertainties affect the derivation of the
scaling relations and overall results of our study.  
In Sections 2 and 3 we describe our observations 
and our imaging data reduction process. 
Catalogs of photometric
and structural parameters for RX J0152.7-1357, RX J1226.9+3332, and 
RX J0142.0+2131 are presented in Section 4.  We discuss internal and external 
consistency of our results and implications on the measurements of the 
FP in Section 5.
A summary is presented in Section 6.
We assume $H_{0} = 70$ km s$^{-1}$ Mpc$^{-1}$, $\Omega_{m} = 0.3$, and 
$\Omega_{\Lambda} = 0.7$.

\section{Observations}\label{obs}

%
We have acquired ground-based imaging for the clusters in our sample 
using GMOS on the Gemini-North 8m telescope.  GMOS spectroscopy
was obtained for $40-50$ galaxies in each cluster field, $20-30$ of which 
turned out to be 
cluster members.  GMOS is described in \citet{hook}. 
Details of how the spectroscopic
sample was chosen and the 
observations and reduction of these data
are described elsewhere \citep[J{\o}rgensen et al. in prep;][]{jorg05,barr05}.  
For each of our clusters we have obtained new ACS imaging 
or use archival HST data 
to study the galaxy structural parameters with high resolution data having
FWHM $= 0^{\prime\prime}.1$.  Data for the three clusters described in this paper were
obtained using the ACS Wide Field Channel.  The ACS/WFC detector consists of
two 2048x4096 chips separated by a gap of 2.5 arcsec with a mean pixel scale $0^{\prime\prime}.049$ 
and a $3^{\prime}.36 \times 3^{\prime}.36$ field of view \citep{sir05}.  

For RX J0152.7-1357 (z=0.83), we used archive data from Program ID 9290 
which includes 4 mosaicked ACS fields observed in 3 filters each, F850LP, F775W, and
F625W, corresponding to the $z',i'$, and $r'-$bands in our GMOS data.
Total exposure times for each field were 4800s in the $i'-$band and
4750s in the $z'$ and $r'-$bands.   
These data contain all 29 of our spectroscopic sample cluster member galaxies.  

For RX J1226.9+3332 (z=0.89), ACS archival data from Program ID 9033 
included four fields 
in two bands each, F606W and F814W, the latter equivalent to the Cousins 
$I-$band.  Exposure times were $8\times500$s.  These fields contain our original 25 
sample galaxies used in \citet{jorg06,jorg07} along with an additional 87 galaxies 
from an extended sample.

We obtained data for RX J0142.0+2131 (z=0.28) from Cycle 12 Program ID 9770 observed on 
UT 2003 November 01 and 2004 July 03. Two fields were imaged
in a single band, F775W. 
The total exposure time for each position was 4420s.  28 of 30 spectroscopic sample
cluster member galaxies were covered by these fields.
See Table \ref{imagetab} for a summary of all the ACS data used in this paper.

\section{Data Processing}\label{dataproc}

\subsection{HST/ACS Initial Processing and photometry}\label{hstproc}

The three clusters discussed in this paper were used to establish the data 
reduction pipeline.
Although the data are distributed already processed from STScI we chose
to redo the reduction starting with the stacking of flattened images in
order to achieve better cosmic ray removal.  In some cases we also wished
to mask out large reflections or saturated stars before stacking the images. 
We used MULTIDRIZZLE v. 2.3, an STScI distributed PyRAF task\footnote{STSDAS and
PyRAF are products of the Space Telescope Science Institute, which is
operated by AURA for NASA} \citep{multidriz02}, 
which drizzles, distortion corrects, and stacks images all in one.

To obtain better relative shifts between separate exposures at each 
pointing, we initially ran MULTIDRIZZLE
to generate a median combined image for each pointing along with 
individually drizzled images shifted to the reference frame of the 
first image in each set.  The images at this point were aligned simply by 
using the header WCS.
We then ran a subset of the Gemini IRAF package\footnote{IRAF is distributed
by National Optical Astronomy Observatory, which is operated by the
Association of Universities for Research in Astronomy (AURA), Inc., under
cooperative agreement with the National Science Foundation.  The
Gemini IRAF package is distributed by Gemini Observatory, which is
operated by AURA.} task IMCOADD which first detects
stars on the cosmic ray-free median image using DAOPHOT.  It then
locates these stars on the individual drizzled images and measures any
additional relative shifts.
This step created a shift file suitable as input 
for MULTIDRIZZLE which was subsequently run a second time to create 
the final stacked, distortion corrected, cosmic ray cleaned image.  
A wrapper script was used to perform the 
necessary steps which included an option to allow manual sky subtraction,
for use when bright stars or reflections effected the image.  
We chose not to perform any resampling during the drizzle 
and we used the default square drizzle
kernel as no improvement was observed with other kernels.

MULTIDRIZZLE output includes weight and context images. 
Each pixel within the context image is encoded with information about
which specific individual images were used during the combining
operation for that pixel (ie. were free from chip defects and CRs). 
We use the context image to generate a map which more generally contains
information about the number of pixels that were combined to create 
each stacked pixel 
value.  This is used to calculate more rigorously the sigma noise map
for each ACS image taking into account the correlated noise produced
by the drizzling \citep{cas00}.  The noise at each pixel is
calculated as
$\sigma^2 = F_A ((RN * \sqrt{N_{comb}} / gain_{eff})^2 + (N / gain_{eff}))$
where RN is the readnoise, N$_{comb}$ is the number of pixels used in the
drizzling, gain$_{eff}$ is the effective gain at each pixel incorporating N$_{comb}$, 
and N denotes the counts in that pixel. F$_{A}$ corrects for the correlated
noise and comes from appendix 6 in
\citet{cas00}.  It depends on the output pixel scale (in this case the fractional 
size compared to the original) and 
the drop size for the drizzling.  In our case, scale and pixfrac are both 1.0  
since we do not drizzle to smaller scales. F$_{A}$ is also dependent on the size of the
area considered since correlated noise scales from single pixel noise
differently than uncorrelated noise.  We did not take this area correction into account
here.  Thus, 
our noise map may incorrectly estimate the true photometric errors.  Although
this would have an effect on the GALFIT error estimates on the individual 
parameters, we do not expect this to affect our results since we
determine parameter measurement uncertainties independently through galaxy simulations
(Section \ref{tests}).

The weight image uses weighting based on exposure times.  To improve
extraction of galaxies in later processing using SExtractor,
we adjust this weight image using our n-pixel map described above to set a very low
weight near image borders and in gap regions where few overlapping
images produce greater noise.

The zeropoint calibration in the AB system is provided by 
STScI for the ACS camera in each filter used \citep{sir05}.
We correct the WCS in the header of the stacked images, which 
were in most cases off by a shift of about $2^{\prime\prime}$ in RA
and $1^{\prime\prime}$ in DEC, 
by using the IRAF tasks CCMAP and CCFIND along with coordinates for
these galaxies previously obtained from our GMOS images 
relative to the USNO catalog \citep{monet98}.  Final coordinates are generally good to 
about $0.2^{\prime\prime}$ in both RA and DEC.

\subsection{SExtractor Object Detection and Photometry}\label{obphot}

We used SExtractor \citep{ba96} v.2.3.2 
to detect all objects in the images having at least 9 contiguous pixels
greater than $2\sigma$ above the sky noise.  
The weight image described above was used
to minimize the number of spurious noise detections predominently
found near image borders and, in some cases, chip gap regions of the combined 
dithered images.

SExtractor was run in association matching mode to specifically recover 
our spectroscopic sample galaxies.  It was then run a second time to
detect all other galaxies in the field, using dual image mode in cases where
fields had been observed in multiple bands.  In this mode, we use the band closest 
to the $i^{\prime}-$band for the initial detection of the objects.  After
using the IRAF task imshift to precisely align images in other bands
to the  $i^{\prime}-$band image, we run SExtractor to
perform photometry in these other bands using the
$i^{\prime}-$band determined apertures.  
This produces matched catalogs of objects in all bands.

After detecting all objects in a first pass through the data, SExtractor
makes a second pass to deblend merged objects.
We ran a number of tests to determine the best
set of SExtractor deblending parameters, DEBLEND\_NTHRESH (sets the number of
deblending levels) and DEBLEND\_MINCONT (minimum fraction of 
pixels a branch must have to be considered a separate object), which would
correctly split and detect objects in our fields.
Aperture check-images were used to determine whether objects
had been detected correctly.  For our spectroscopic
sample galaxies, the best set was found
to be DEBLEND\_NTHRESH = 32 and DEBLEND\_MINCONT = 0.01 as these
parameters best recovered galaxies in their entirety
without excess splitting while still correctly separating
neighboring objects.
While this worked for most program galaxies in all clusters,
there were cases of galaxies with a lot of structure that were split too
much, or galaxies with foreground stars which required
greater deblending. For these galaxies we adjusted the
DEBLEND\_MINCONT slightly lower or higher as necessary until
that individual galaxy was verified by eye to have been detected correctly.
For the non-sample galaxies in the images, many consisting of late
type, irregular galaxies, we used a compromise set of parameters of
DEBLEND\_NTHRESH = 16 and DEBLEND\_MINCONT = 0.01.
This minimized oversplitting while still separating
most neighboring objects.
 
The SExtractor output, which we use primarily as input for
more rigorous galaxy profile fitting, included various
magnitude (and flux) determinations, fractional pixel centroids, object class, 
ellipticity,
position angle, and various moment and size measurements. We take
the parameter $m_{best}$, a combination of adapted aperture and
corrected isophotal magnitudes, as the estimate for total magnitude.
Measured parameters for objects detected multiple times in overlapping
image frames were averaged together.

\subsection{GALFIT Surface Photometry}\label{fitting}

To perform surface brightness profile fitting, we used GALFIT, a 
two-dimensional profile fitting program written by \citet{peng02}.
GALFIT has the advantage over other similar profile fitting codes because it
simultaneously fits neighboring galaxies. This improves the fit for 
r$^{1/4}$ law galaxies which contain a significant amount of light in the outer
wings of the profile \citep{hmb07}.  GALFIT provides the user with
a choice of analytic surface brightness profile functions with which to fit the 
galaxies including Sersic \citep{sers}, de Vaucouleurs (r$^{1/4}$ law), exponential 
disk, bulge/disk deconvolution, and psf.  

The user must provide
initial guesses for the profile parameters and a PSF template 
used to deconvolve the image PSF during fitting.    We
use the SExtractor output m$_{best}$ for total magnitude, x, y, ellipticity,
theta (corrected to GALFIT orientation), and flux radius
(set within SExtractor to be an estimate of the half-light radius and
converted to a$_{eff}$ for GALFIT) as input for GALFIT.
Tests were made using simulated galaxies to determine the most
appropriate PSF model and size.  See
Section \ref{models} for more details of this testing process.  The chosen PSFs were
created by first running Tiny Tim \citep{tt95} to generate
raw distorted ACS model PSFs for a grid of locations spread uniformly over
the two ACS chips.  These were then added to a set of blank ACS images with the
appropriate shifts
and multidrizzled to produce PSFs combined in the same manner as the real data.  
These PSFs were re-extracted from the rectified frame as individual 
PSF templates for use with GALFIT.
We use a 1-time sampled $9^{\prime\prime}$ PSF for the GALFIT fitting procedure.

We fit all galaxies with both r$^{1/4}$ law and Sersic \citep{sers}
profiles. The Sersic profile has the form

\begin{equation}
\Sigma(r) = \Sigma_{e} \exp[-\kappa((r/r_{e})^{1/n} - 1)]
\end{equation}

\noindent
where r$_{e}$ is the half-light radius, $\Sigma$ is the surface 
brightness, and n is the Sersic index.  $\kappa$ is coupled to
n to ensure that half of the total flux lies within r$_{e}$.
The total flux can be calculated by integrating out to r $= \infty$:
\begin{equation}
F_{tot} = 2 \pi r^{2}_{e} \Sigma_{e} e^{\kappa} n \kappa^{-2n} \Gamma(2n)q /R(c)
\end{equation}
\begin{equation}
R(c) = \pi(c+2) / (4\beta(1/(c+2),1+1/(c+2)))
\end{equation}

\noindent
\citep{peng01} where q is the axial ratio, b/a, c is the diskiness-boxiness parameter,
and $\beta$ is the Beta function. 
The advantage of fitting with a Sersic profile is that the
exact form of the profile need not be known a priori 
since the index, n, can vary to fit the full range of 
possible galaxy profiles
from n $= 1$ exponential disks to n $= 4$ ellipticals/spheroidal
components. 

Since ACS images are large, we used a 
wrapper script which first extracted a smaller panel for each galaxy 
from the image, with a panel
size determined to be $\sim25\times$r$_{e}$ of the galaxy to be fitted, having a
minimum size of 250 pixels and maximum of 1000 pixels on a side.
Because light from the extended profiles of neighboring objects 
will influence the profile fitting of the primary galaxy,  
all galaxies within the panel were 
either fitted or masked.  Masking was necessary 
to limit the number of galaxies fit and speed up the execution of the fitting.
We therefore simply masked out all objects detected by SExtractor more than 
four magnitudes below the mean sample galaxy brightness. 
All other objects in the panel were fitted with Sersic profiles. 
These image sections were also
checked by eye to ensure that no objects had been missed by SExtractor.

The wrapper script fit all cluster sample galaxies with both a Sersic 
and a de Vaucouleurs (r$^{1/4}$ law) profile.  For each profile we have a 
minimum of 6 free parameters, fitting for 
total magnitude, effective radius, ellipticity,
position angle, and x,y position.  In the case of the Sersic profile,
the Sersic n was set initially at 1.5 and allowed to vary.
After fits with both Sersic and r$^{1/4}$ law profiles were performed, any object
best fit with n $> 3.0$, was refit with a Sersic profile using n=4.0 
along with the output from the r$^{1/4}$ fitting as initial guesses.  All
parameters were allowed to vary.  We included this iteration because the initial 
guesses 
tended to influence the fit and SExtractor did not always produce the
most accurate estimates for input. 
Neighbors were fit simultaneously with Sersic profiles while the sky value
was held constant at the value provided by SExtractor.  
Best fits are determined by minimizing the $\chi^{2}$ residual, adjusting
all free parameters simultaneously.

\begin{figure}[t]
\begin{centering}
\plotone{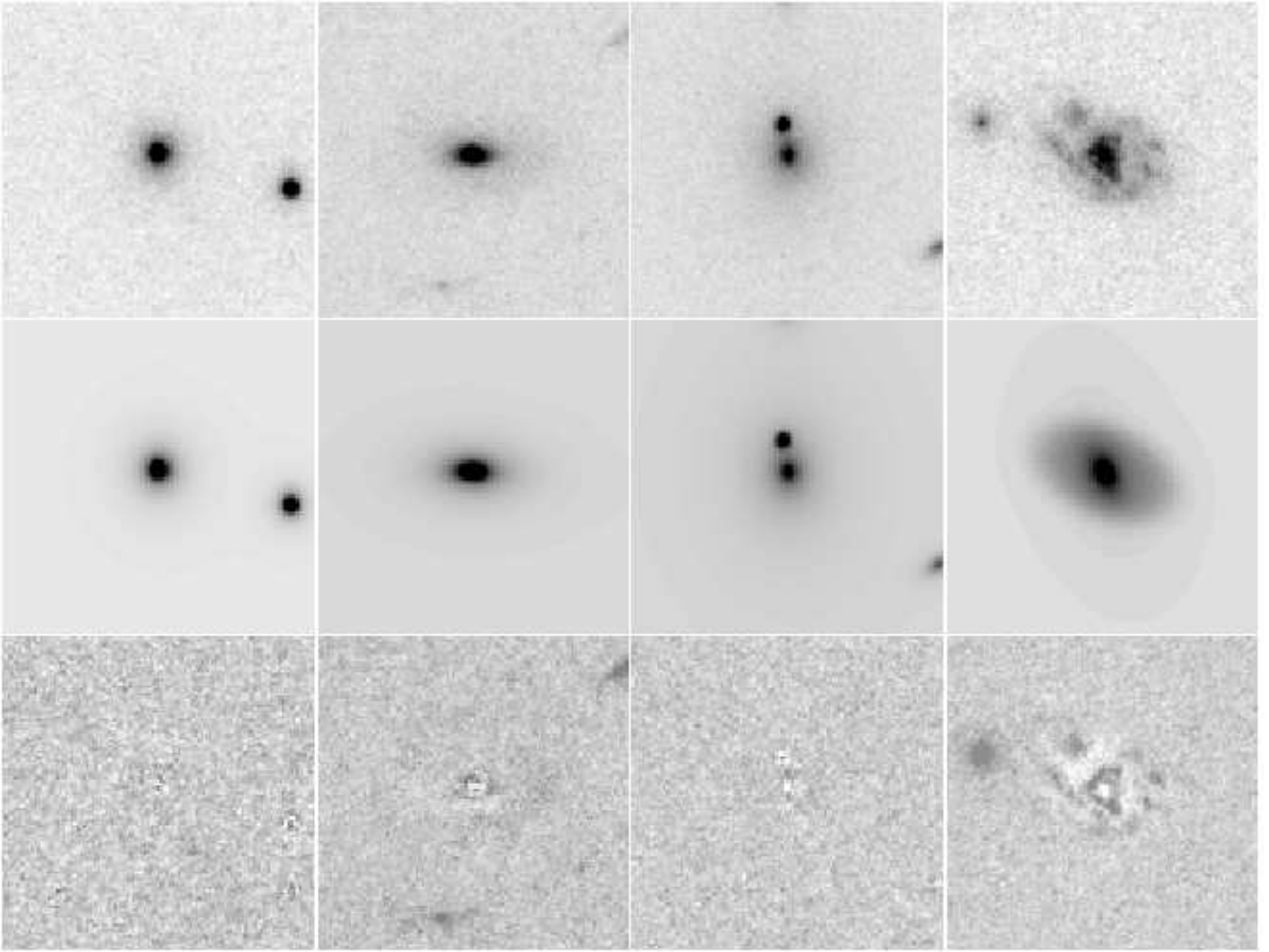}
\caption[Examples of profile fits]{Several examples of output
images produced by GALFIT. These include, from top to bottom, the panel from the
original image, a model image with the primary galaxy fit with an r$^{1/4}$ profile,
and the residuals from the best fit.
In every case, the primary galaxy being fitted is the central object.
Galaxies are from RX J0152.7-1357: 737 (with Sersic fit n = 3.71), 643 (n = 3.58),
1567 (n = 3.17), 1385 (n = 1.11).  Image sections are 5.2 arcsec across.
\label{galexmpl}}
\end{centering}
\end{figure}

\begin{figure}[t]
\begin{centering}
\plotone{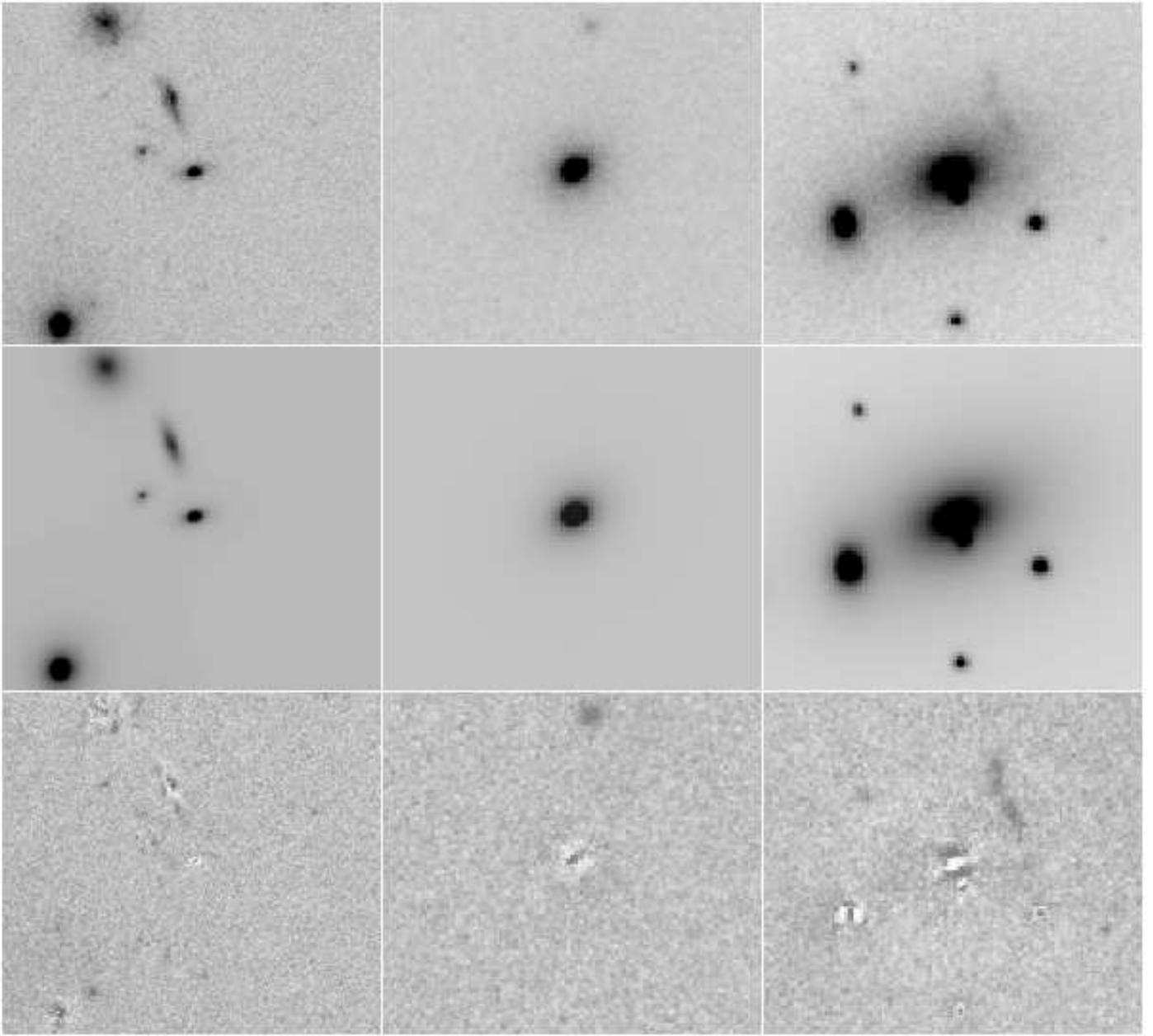}
\caption[Examples of profile fits]{Best r$^{1/4}$ law fits as in Figure \ref{galexmpl}.
Galaxies are from RX J1226.9+3332. From left to right: 1025 (with Sersic fit n = 6.96), 529 (n = 4.74),
and 563 (n = 3.92). Image sections are 11.8, 5.9, and 5.9 arcsec across respectively.
\label{galexmpl2}}
\end{centering}
\end{figure}

\begin{figure}[t]
\begin{centering}
\plotone{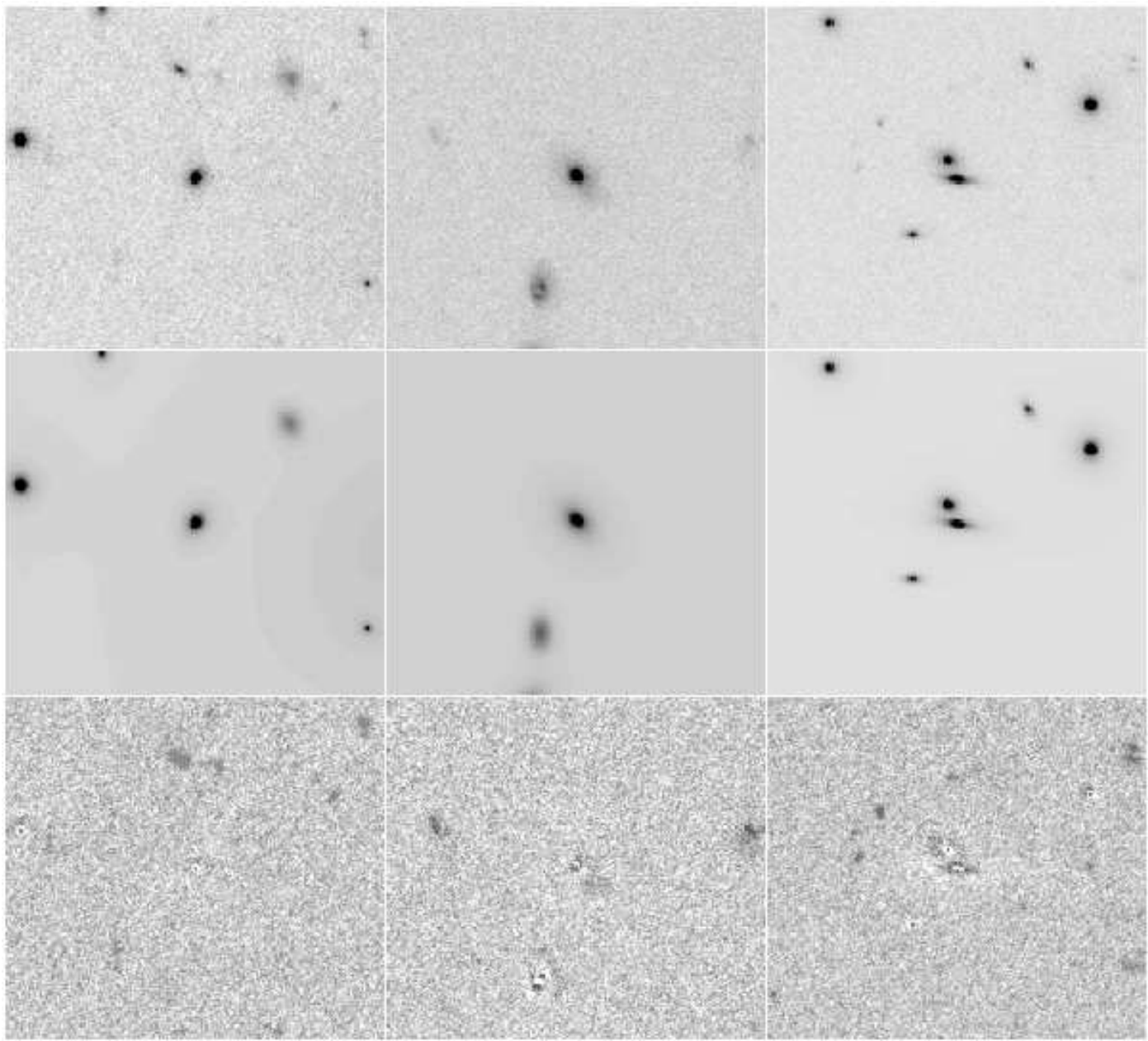}
\caption[Examples of profile fits]{Best r$^{1/4}$ law fits as in Figure \ref{galexmpl}.
A further 3 examples from RX J1226.9+3332: 630 (with Sersic fit n = 3.29), 739 (n = 3.14), and
1251 (n = 1.90). Image sections are 11.8 arcsec across.
\label{galexmpl3}}
\end{centering}
\end{figure}

Final parameters produced by GALFIT include total magnitude,
effective radius, axial ratio, position
angle, x and y centroids, Sersic n, and a $\chi_{\nu}^{2}$ value of the fit,
along with associated uncertainties for all the derived parameters.
Output also consisted of a multiple extension image including the 
original panel, a model image, and the residuals from the fit.  
We display several examples of r$^{1/4}$ law fits in Figures \ref{galexmpl} - \ref{galexmpl3}.
Larger residuals are evident in galaxies 
with best fit Sersic index n closer to 1 and in those exhibiting
features beyond a smooth profile.
Final $\chi^{2}_{\nu}$ values typically indicated good fits.
Median values of $\chi^{2}_{\nu}$ for fits with both profiles in all three clusters were
$\sim 1.4$ with maximum values for a few galaxies between $4 - 8$.
These were cases of galaxies with structures that were not well modeled with 
single profiles or of galaxies with problematic neighbors.

GALFIT quoted uncertainties generally underestimate the true uncertainties of 
the parameter measurements
since these assume the model is a perfect representation of the real galaxy.
We list the GALFIT median measurement uncertainties in Table \ref{mederr}.
Uncertainties for the low redshift cluster were slightly lower than for
the two high redshift clusters.
For all three clusters, uncertainties for r$^{1/4}$ law parameters were slightly 
lower than for the Sersic
fits. However, this does not imply that these fits were necessarily better and could
indicate that these models were a poorer representation of the real data.   
In cases where a model is a poor match to the data, 
statistical uncertainties can be smaller because
a small change in the parameters will produce a large change in 
$\chi^{2}$ \citep{peng01}.  
The Sersic function with its extra free parameter is better able to fit a
galaxy profile that deviates slightly from the traditional r$^{1/4}$ law.
While the uncertainties determined from profile fitting may be 
statistically accurate, they do not provide realistic uncertainties for the 
various derived parameter values.
We describe in the Section \ref{models} our use of simulated galaxies to derive more
realistic measurement uncertainties.

\section{Results}\label{cata}

Our high redshift cluster galaxies have half-light radii for our assumed
cosmology ranging
in size from $\sim 0.8$ to 35 kpc with magnitudes between  $20.3 < i^{\prime} < 23.7$.
The full RX J0152.7-1357 spectroscopic sample includes 41 galaxies, 36 of which are contained within
the ACS images.  Structural parameters have been obtained for all 29 spectroscopically 
confirmed members.
The RX J1226.9+3332 sample of 112 galaxies with measured parameters includes the 
original spectroscopic sample along with 63 galaxies having archival spectroscopy. 
54 galaxies in this full sample are members. 
The 28 galaxies in the RX J0142.0+2131 sample have magnitudes ranging 
from $16.9 < i^{\prime} < 22.0$ and sizes
$0.7 < r_{e} < 34$ kpc.  In Tables \ref{rxj0152cat} - \ref{rxj0142cat},
also available electronically,
we provide the measured and derived photometric and structural
parameters for these galaxies.
Galaxies are sorted by RA and columns are as follows:

\noindent
Column (1) Galaxy ID.  This number comes from our original
SExtractor detection in GMOS images.  We use this number to identify
the galaxies in all our publications.

\noindent
Column (2) RA (J2000.0)

\noindent
Column (3) DEC (J2000.0)

\noindent
Column (4) Number of measurements made from different images for each galaxy.
Table entries are average values from all measurements.

\noindent
Column (5) Total apparent magnitude derived from r$^{1/4}$ law profile fits.  This
is measured in the $i^{\prime}-$band for RX J0152.7-1357 and RX J0142.0+2131,
and $I-$band for RX J1226.9+3332.  Magnitudes are uncorrected for reddening.

\noindent
Column (6) $\log$ (r$_{e}$) from our r$^{1/4}$ law fits, with r$_{e}$ in arcsec.  We take
the radius as the geometric mean of the semi-major and minor axes, r $= (a b)^{0.5}$.  

\noindent
Column (7) Mean surface brightness within the effective radius derived from 
total magnitude and r$_{e}$ from our r$^{1/4}$ law fits, in mag arcsec$^{-2}$, using 
$\langle \mu \rangle_{e} = m_{tot} + 2.5 \log 2 \pi + 2.5 \log r_{e}^{2}$.

\noindent
Column (8) Total magnitude from Sersic function fits, bands as in Column (5).

\noindent
Column (9) $\log$ (r$_{e}$) from Sersic function fits, with r$_{e}$ in arcsec.

\noindent
Column (10) $\langle \mu \rangle_{e}$ derived from total magnitude and r$_{e}$ from 
our Sersic function fits, in mag arcsec$^{-2}$.

\noindent
Column (11) Best fitting Sersic function parameter, n.

\noindent
Column (12) Position angle, North through East.

\noindent
Column (13) Ellipticity, derived from the fitted axial ratio.

\noindent 
Column (14) Cluster members are denoted by a '1', non-members
by '0'.  Not every galaxy in our sample proved to be a cluster member, but 
all sample galaxies for which we were able to measure structural parameters 
are included in these tables.  A blank entry indicates no redshift is available.
These include galaxy
ID 910, required fitting as a neighbor of galaxy ID 899,
galaxy IDs 1009 and 1253 which had archive spectroscopy but with too low S/N to derive a 
redshift, and galaxy ID 1254 for which we were unable to extract a
redshift due to confusion with a second object in the slit.

\section{Discussion}\label{tests}

\subsection{Internal consistency}\label{models}

Simulated galaxies are used to investigate sources of error in
our pipeline and test the accuracy of the software used. GALFIT,
for example, produces uncertainties for the output parameters which
tend to be too low since these are strictly
random uncertainties and do not take into account the fact that the
profile we fit each galaxy with may not be an exact match to the true
galaxy profile \citep{peng01}.  We therefore use the simulated galaxies to 
determine more realistic measurement uncertainties and to test how these uncertainties
affect our FP measurements.  The simulated galaxies are also used
to test the effect of different PSF models and sizes for convolution in the 
galaxy profile fitting.  In this section, we describe the methods used to 
simulate the galaxies, describe tests to determine the best PSF to use 
for the 2-D surface brightness profile fitting, and determine expected 
uncertainties in our galaxy parameter measurements.

\subsubsection{Galaxy simulations}

To generate realistic galaxies, we used the structural parameters of 
148 galaxies in our low redshift Coma cluster comparison
sample \citep{jf94}. We transformed the values
of M$_B$ to the filter and redshift of each cluster using stellar population
models of \citet{bc03}. 
For simulating galaxies at z $= 0.83$, we transform the M$_B$ magnitudes of Coma cluster galaxies
to the observed $i^{\prime}$ band using 
\begin{equation}
M_B = B_{rest} - DM(z), 
\end{equation}
\begin{equation}
B_{rest} = i^{\prime} + 0.8026 - 0.4268(i^{\prime} - z^{\prime}) - 0.0941(i^{\prime} - z^{\prime})^2, 
\end{equation}
\citep{jorg05,jorg07}
and a mean empirical color, $(i^{\prime} - z^{\prime}) = 0.8$,
for our early type galaxies at this redshift.
For z $= 0.28$, we use
\begin{equation}
B_{rest} = i^{\prime} + 0.4753 + 1.6421(r^{\prime} - i^{\prime}) - 0.0253(r^{\prime} - i^{\prime})^2
\end{equation}
\citep{barr05}
with a mean empirical color $(r^{\prime} - i^{\prime}) = 0.5$.
We do not include a separate simulation for z $= 0.89$ as those made for z $= 0.83$ 
are considered representative.
We scale r$_{eff}$ to the appropriate redshift within our assumed cosmology.  
To create a large sample of simulated galaxies we added a small amount of 
random scatter to the magnitudes, r$_{eff}$, and Coma galaxy axial ratios 
while randomly generating positions and position angles.  
The Sersic parameter, n, was allowed to vary randomly between 0.5 - 5.5,
except for a few sets of simulations where values ranged between 3.5 - 4.5
in order to increase the total number of early type galaxy simulations.
The diskiness/boxiness of each galaxy was also allowed to vary randomly.
Galaxies were added as perfect Sersic profiles, albeit with 
Poisson noise added to the galaxy images. 

In order to simulate galaxies with multiple components, we also
generated a set of bulge + disk galaxies.  In these cases, bulges
were created by modeling from the Coma galaxy structural properties.  Disks
were then added to the same position assuming disk-to-total light ratios 
ranging randomly from 0.25 - 0.65 with disk sizes ranging from (1 - 1.4) r$_{e(bulge)}$.
The inclination was varied from 0 to 90 degrees,
uniformly in $\cos i$. 
As in \citet{jf94}, we assign each component an intrinsic
axial ratio: b/a(bulge) = 0.7 and b/a(disk) = 0.15.
These values are chosen such that when transformed into observed values
according to the inclination by
\begin{equation}
b/a_{observed} = \sqrt{1 - (1-(b/a_{intrinsic})^2)(\cos{i})^2}
\end{equation}
\citep{sfs70}, they will display the same range of axial ratios found in 
real galaxies.  

Simulated galaxies were added to real 
images using two different means.
Galaxies were generated with ARTDATA by defining a
Sersic profile and using the Coma based parameters.  They were
also created using GALFIT itself by turning off the fitting for all
parameters thereby
forcing GALFIT to output models of the input parameters.
No differences were observed in the galaxies generated or in the results obtained
from the two methods, but both were used to ensure that
no bias was produced from the simulation method.

Before being added to images, simulated galaxies were first convolved
with a PSF and noise was added to the galaxies with the
IRAF task MKNOISE.  PSFs were created using 4-6 unsaturated real
stars having high S/N using the routines in the
IRAF package DAOPHOT. These were created for each ACS image that simulated galaxies
were added to.

In order to measure
realistic GALFIT parameter uncertainties we generated a total of over $2600$
z $= 0.83$ and $2500$ z $= 0.28$ simulated galaxies.  
So as to not increase the surface density of objects in the images and thereby 
affect recovery of the surface brightness profile parameters, no more than 50 
were added at a time to the real cluster images.
It is important to note that all galaxies are generated as 
perfect Sersic profiles and recovered as
Sersic and r$^{1/4}$ law profiles.  We do not simulate structure, such as
arms, bars, double nuclei, shells, or twisted isophotes, all of which would
effect the fitting.
For the measurement of the cluster FPs, we have
included only non-emission line, non-star forming galaxies
which we take to be the early types. However, at higher redshift,
one might expect to find more late-type features in these non-star forming galaxies.
As a simple test of how added structure will affect single profile model fits, we
use the set of simulated galaxies having both bulge + disk components.
                                                                                                                            
\begin{figure}[t]
\begin{centering}
\includegraphics[angle=0,totalheight=3.5in]{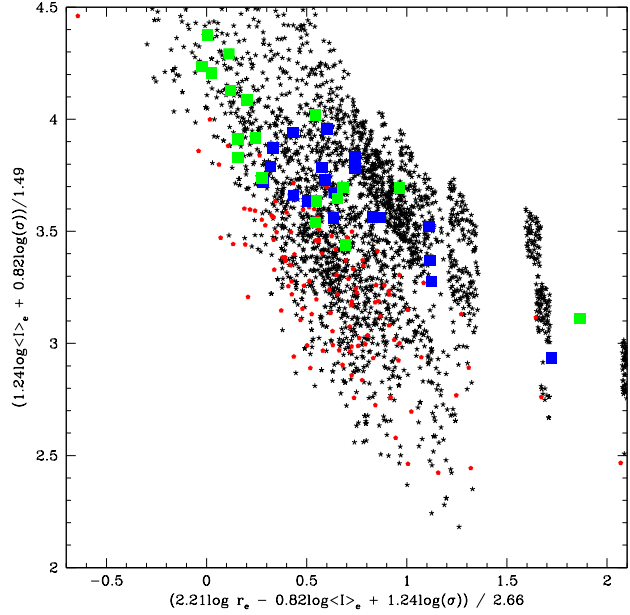}
\caption[False galaxies on the FP]{The face-on FP.
Blue squares represent galaxies of RX J0152.7-1357, green squares RX J1226.9+3332,
and red dots Coma sample galaxies.  Black stars denote the location of
our simulated galaxies in this FP. These simulated galaxies
span the full range of the high z FP.
\label{fprange}}
\end{centering}
\end{figure}

In Figure \ref{fprange}, we display a face-on view of the FP with
our Coma cluster and high redshift galaxies plotted. We overlay our high redshift
simulated galaxy sample.  Expected velocity dispersions given the input
r$_{e}$ and $\langle \mu\rangle_e$ for each simulated galaxy were calculated
using the FP equation for Coma, $\sigma = (\log r_{e} + 0.82 \log \langle I\rangle_{e} + 0.443)/1.3$
\citep{jorg06}.
It can be seen that our simulated galaxies span the same region of the FP
as the real galaxies.   

\subsubsection{Choosing the best PSF for profile fitting}

In order to determine the most appropriate PSF to use for deconvolution during the surface 
brightness profile fitting, we use a subset of 700 simulated galaxies to test the recovery of parameters
with GALFIT using several different PSF models and sizes.
Images were run through our reduction pipeline starting with SExtractor object detection
in order to provide initial guesses for the GALFIT profile fitting.  
We tested the GALFIT recovery using four different PSFs: the PSF based 
on real stars, a raw $3^{\prime\prime}$ Tiny Tim PSF generated for the 
appropriate location on the ACS chip, and $3^{\prime\prime}$ and 
$9^{\prime\prime}$ drizzled PSFs (see Section \ref{fitting} for details).  
Galaxies were fitted with both
Sersic and r$^{1/4}$ law profiles.  The r$^{1/4}$ law fits show
greater discrepancy from input values than Sersic function fits because 
galaxies were added with a Sersic parameter n which varied randomly
between 1 and 4.5 for these tests.  

In Table \ref{psfresults}
we list the average differences and standard deviation between 
input and recovered magnitude, effective radius (in arcsec),
and the combination of parameters that enter into the FP,
$\log r_e - \beta \log \langle I \rangle_e$, hereafter referred to as the 
Fundamental Plane parameter (FPP), where $\langle I \rangle_e$
is the surface brightness within $r_e$ in units of $L_\odot$ pc$^{-2}$ and
$\beta \sim -0.8$ \citep{jorg06}.
The average difference between input and recovered
values is nearly identical regardless of the PSF used. However, we do find that
using the raw Tiny Tim PSF consistently provides the worst results.  
Using $9^{\prime\prime}$ drizzled PSFs showed very slight improvement
over $3^{\prime\prime}$ drizzled PSFs and very similar results to the
real PSFs.  
A test using a subsampled PSF showed no improvement in the output results.
Due to the difficulties producing a real PSF from the few stars in 
our images, we therefore chose to use the model $9^{\prime\prime}$ drizzled
PSF. 
In the event that the Tiny Tim modeled $9^{\prime\prime}$ drizzled PSF is not the
best representation
of the true PSF, we note that the differences in the FPP due to the
PSF used are insignificantly small.

\subsubsection{Structural parameter measurement uncertainties}

The full set of simulated galaxies were used to investigate realistic uncertainties in
structural parameter measurements.  Average offsets (recovered - input values)
and the associated rms scatter in these differences are provided in
Tables \ref{devresults} and \ref{fppresults}. 
As with the PSF tests, simulated galaxies were recovered using the same methods as real galaxies.
The $9^{\prime\prime}$ drizzled PSF described in the previous section is used for the 
surface brightness profile fitting of both real and simulated galaxies.

\begin{figure}[t]
\begin{centering}
\includegraphics[angle=0,totalheight=4.0in]{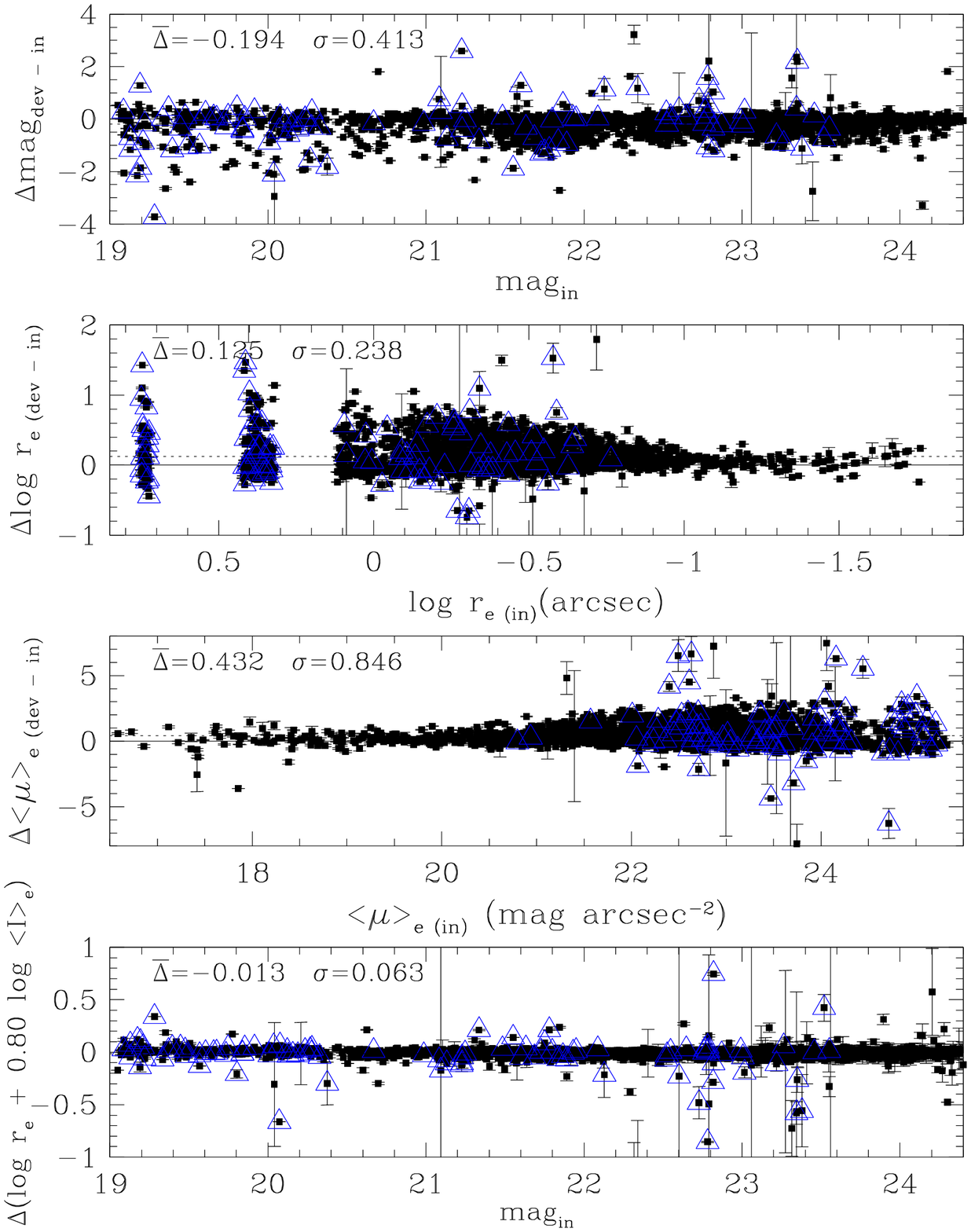}
\includegraphics[angle=0,totalheight=4.0in]{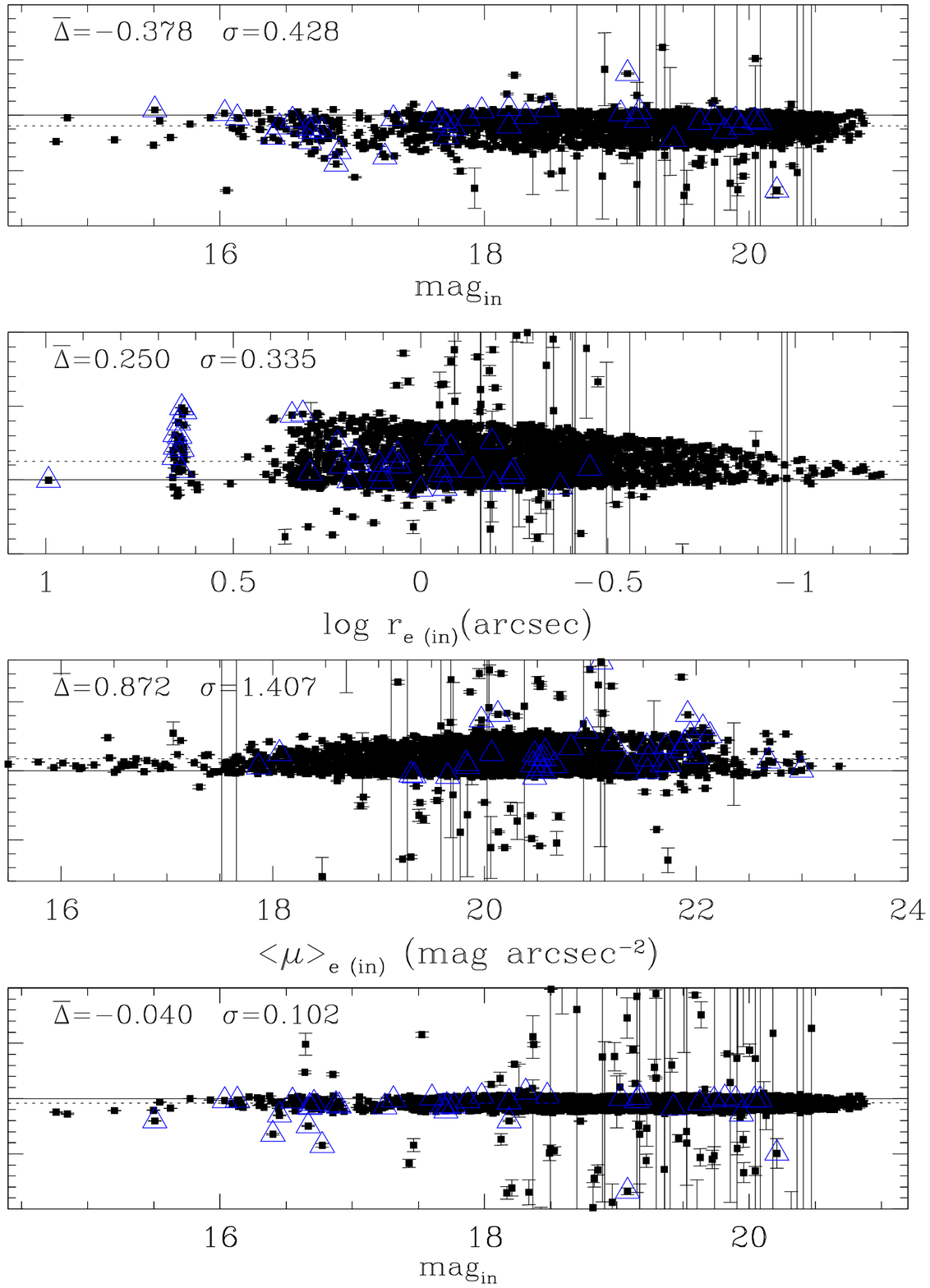}
\caption[Simulation results for de Vaucouleurs profile]{
Left: z = 0.83 simulated galaxy set.  Right: z = 0.28 set.
From fits with r$^{1/4}$ law profiles, we display
the difference in magnitude,
r$_{e}$, $\langle \mu\rangle_{e}$, and the FP parameter 
($\log r_{e} - \beta \log \langle I\rangle_{e}$, see text) 
between input and recovered values.  Points enclosed by triangles
have neighbors within 1 r$_{e}$ of the primary galaxy being fit.  
\label{alldevres}}
\end{centering}
\end{figure}

We compare the r$^{1/4}$ law recovered parameters from GALFIT with the input
values for all simulated galaxies in Figure \ref{alldevres}. 
We find small average systematic offsets and associated rms scatter
from the true magnitude and $\log$ r$_e$ (arcsec) of $-0.2\pm0.4$ 
and $0.1\pm0.2$ for the high z sample.
For the lower redshift set, we find offsets of
$-0.4\pm0.4$ and $0.3\pm0.3$ respectively.  
Average measurement offsets and rms scatter for the FPP are 
only $-0.01\pm0.06$ and $-0.04\pm0.10$ for the two samples.

\begin{figure}[t]
\begin{centering}
\includegraphics[angle=270,totalheight=2.5in]{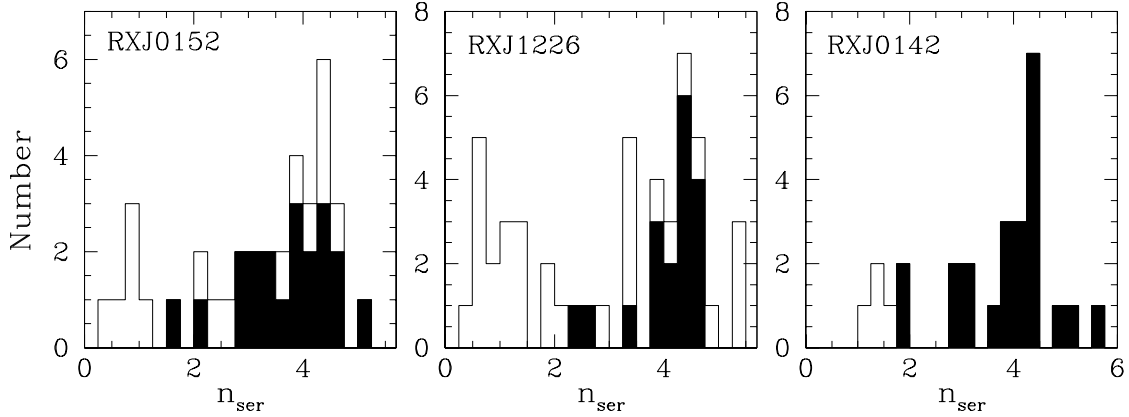}
\caption[Range of values for Sersic parameter n]{The open histogram 
displays the range of measured Sersic parameter n values found for
our real galaxies in clusters RX J0152.7-1357, RX J1226.9+3332, and RX J0142.0+2131. 
The solid histogram
shows the Sersic n values for only those galaxies which we use to
study the cluster FP.
\label{nhist}}
\end{centering}
\end{figure}

\begin{figure}[t]
\begin{centering}
\includegraphics[angle=0,totalheight=4.0in]{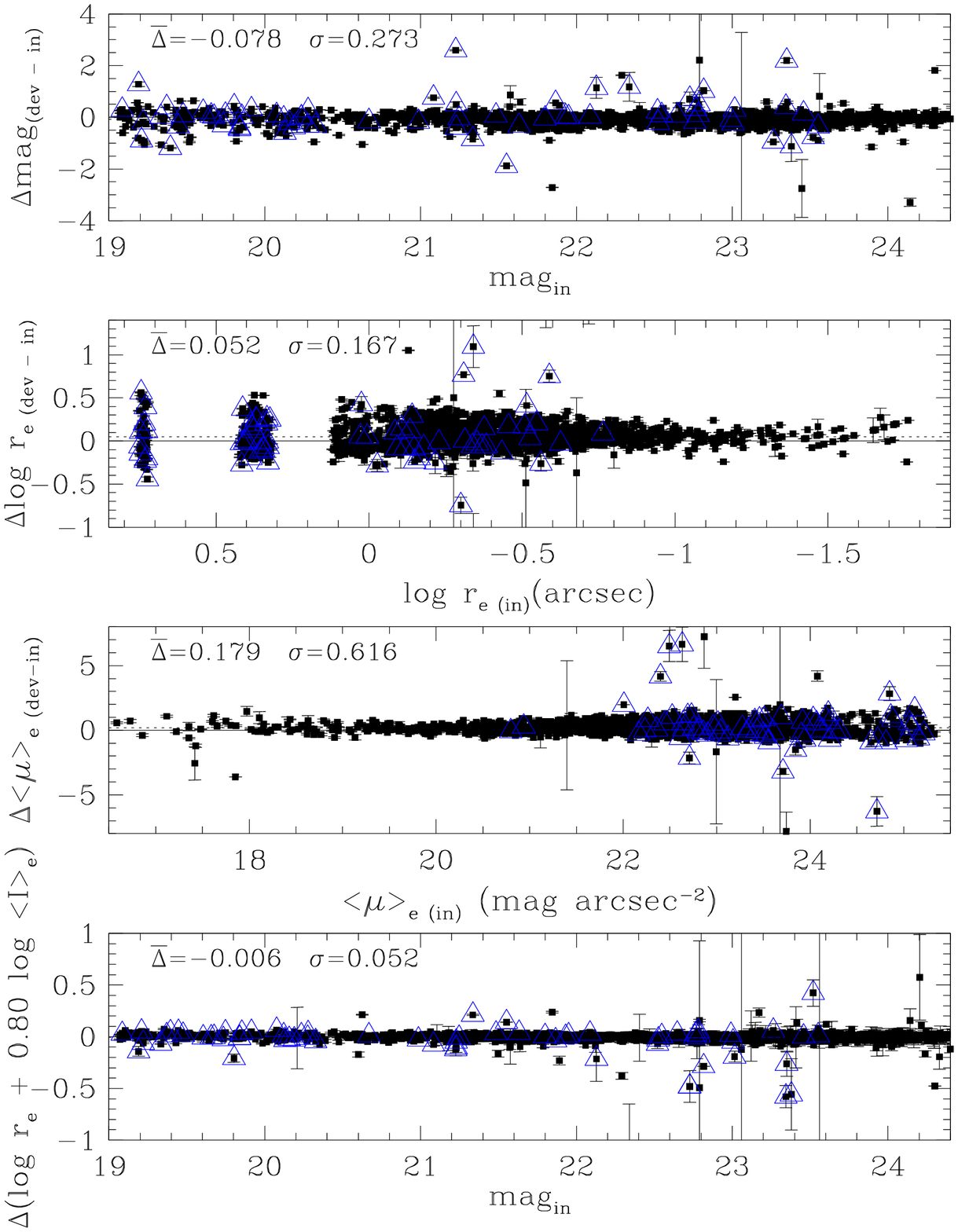}
\includegraphics[angle=0,totalheight=4.0in]{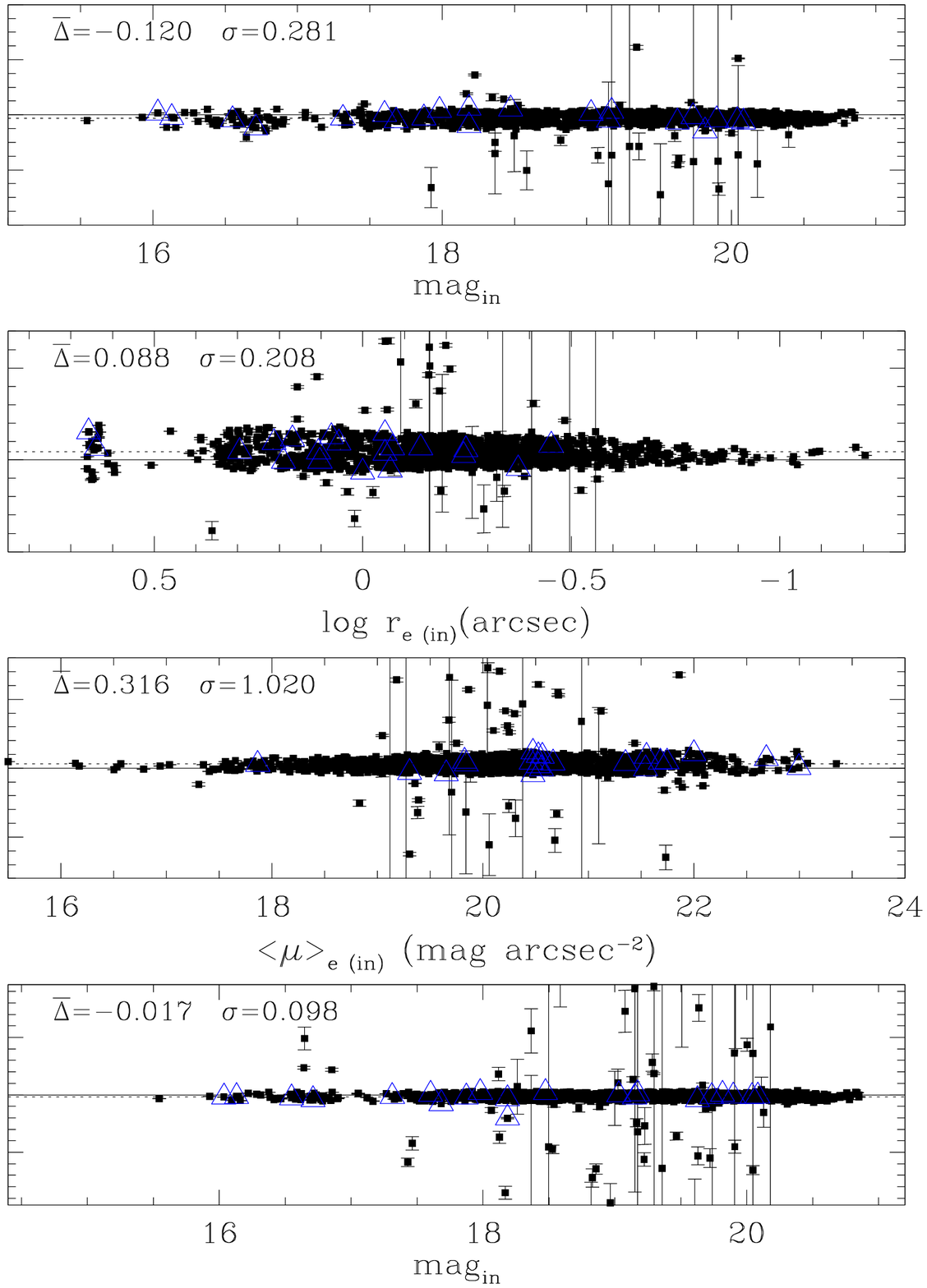}
\caption[Simulation results for n > 2]{Differences
between input parameters and values recovered from r$^{1/4}$ law profile fitting,
but for only those galaxies created with n $> 2$ Sersic profiles, within the 
range of the real galaxies used in the FP.
Symbols are the same as in Figure \ref{alldevres}.  The
results for the high redshift sample are on the left, lower redshift sample on 
the right.
\label{devres}}
\end{centering}
\end{figure}

This is for the full set of simulated galaxies including those that were 
created with exponential 
profiles (n $\sim 1$) but which were recovered with n = 4, r$^{1/4}$ law
profiles.  To obtain realistic uncertainties in our FP parameters,
we must compare only the range of parameters from galaxies that went into constructing
our FPs.  We plot a histogram of the 
Sersic n values measured by GALFIT for our real galaxy samples in Figure \ref{nhist}.
The shaded histogram displays only those galaxies used in our FP.  Galaxies
which were not included in our FP either had emission line
spectra, $\log$ Mass $< 10.3$, or n $< 1.5$ \citep{jorg06}.  We therefore
repeat the comparison between input and recovered parameters in Figure \ref{devres} for
galaxies created with n $> 2.0$ profiles.  It can be seen from Table \ref{devresults}
that while magnitude and size errors have decreased, the systematic
error and random uncertainty in the FPP of $-0.01\pm0.05$ and $-0.02\pm0.10$ 
for the two samples 
exhibit little change.  The larger rms scatter for the lower redshift sample is due
primarily to a few outliers with failed fits.  In our real galaxy sample all failed
fits are flagged and refitted or left out of the FP measurement. We therefore consider
this measurement uncertainty to be an upper limit.

\begin{figure}[t]
\begin{centering}
\plotone{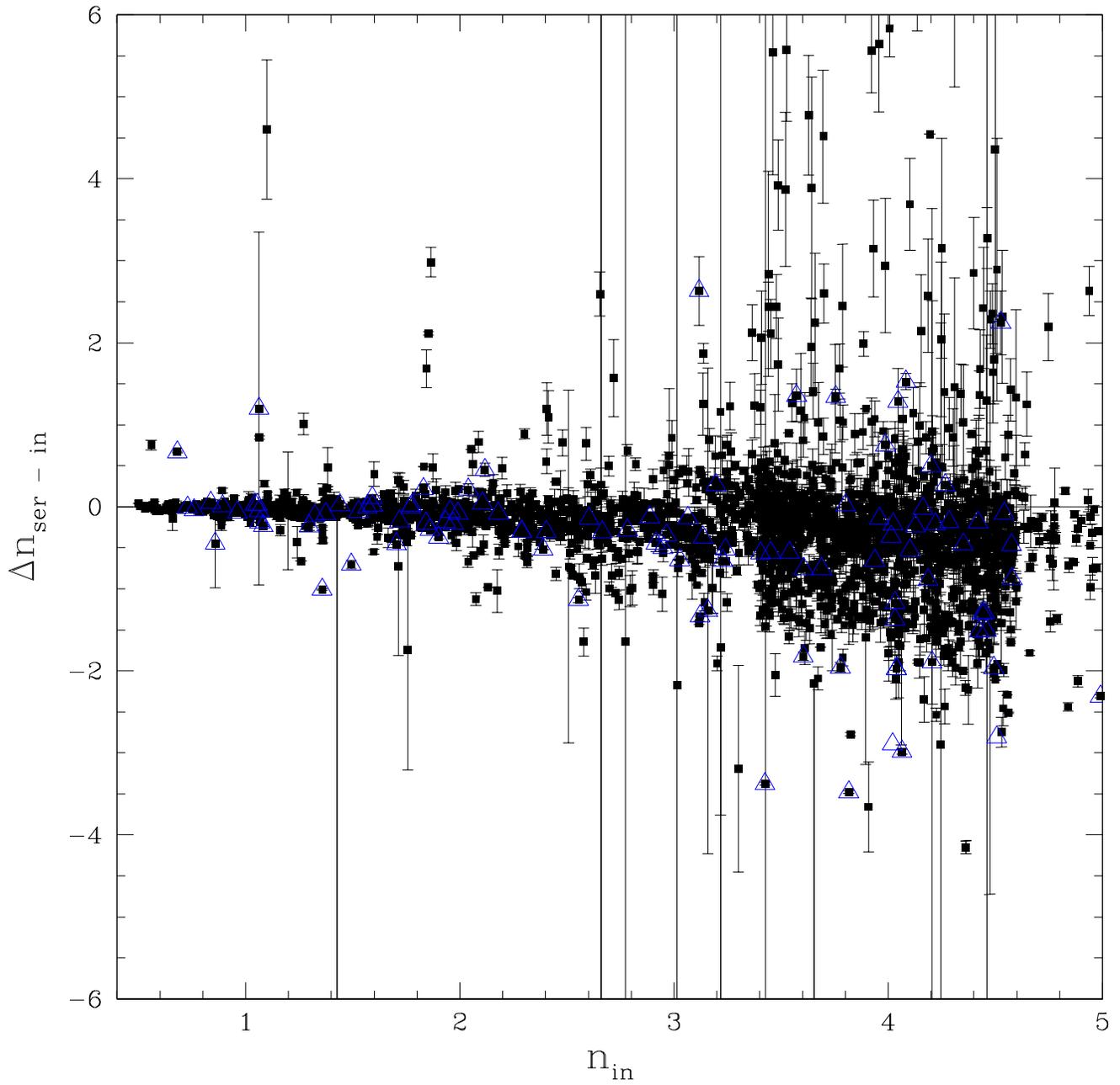}
\caption[Simulation results using Sersic profiles]{Difference in Sersic n
recovered and input values for simulated galaxy profiles fit with 
Sersic functions. 
\label{serres}}
\end{centering}
\end{figure}

Measurement errors are similar when fitting with a Sersic function, allowing n to 
vary.  
Systematic offsets are slightly lower for the magnitude and
r$_{e}$ measurements, but there is no significant improvement in the FPP over fits 
made with r$^{1/4}$ law profiles.  
In Figure \ref{serres}, we display the error
in recovered n as a function of input n.  It is apparent that random error increases
with increasing n.  This is likely because fitting of n=1 type galaxies tends
to be more robust than n=4 galaxies where there is greater weight in the wings of
the galaxy and fits are more easily affected by the sky values and near 
neighbors.  

\begin{figure}[t]
\begin{centering}
\includegraphics[angle=0,totalheight=4.5in]{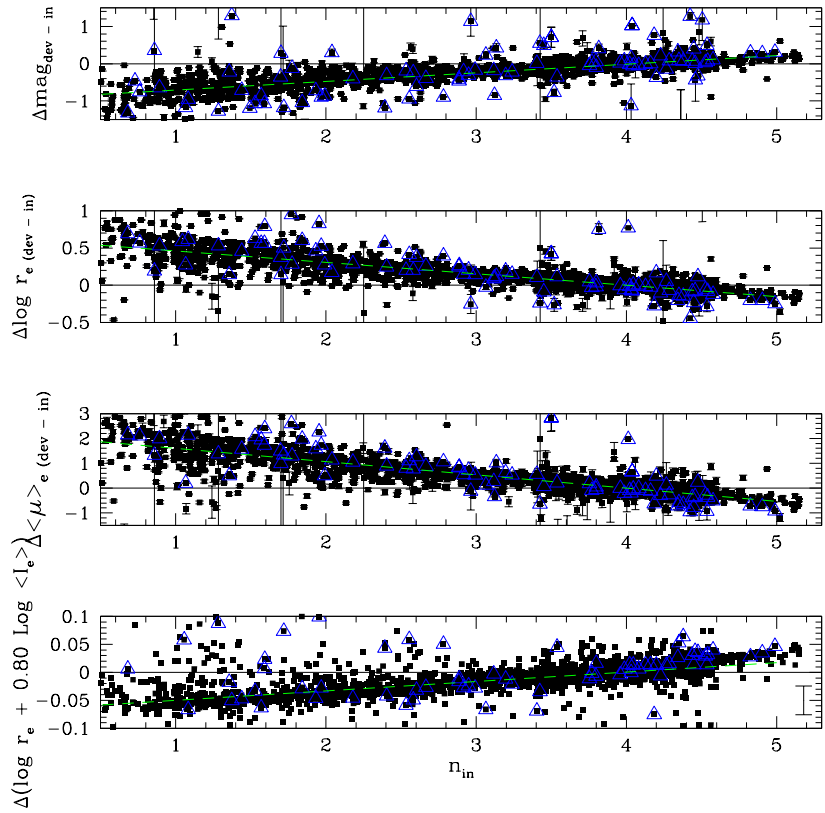}
\caption[Difference as function of n]{
Errors in the  r$^{1/4}$ law measured parameters and derived FPP
as a function of input Sersic parameter n.
For clarity, the errorbar in the lower plot represents the average
measurement uncertainty.
\label{deln}}
\end{centering}
\end{figure}

Because we expect the intrinsic value of n to affect the results for
r$^{1/4}$ law fits, we plot the error in r$^{1/4}$ law recovered 
parameters as a function of input Sersic n in Figure \ref{deln}.  Fluxes and
sizes are overestimated for intrinsically low n galaxies because the r$^{1/4}$ law profile
imposes broader wings than the galaxy has.  Magnitude and half-light radius
are conversely underestimated at
high n ($> 4$). Because flux and size are highly correlated, the errors in the FPP are 
reduced, but can be as large as 0.05 for n = 1
exponential disk galaxies fitted incorrectly with a r$^{1/4}$ law profile.
A linear fit to the offset from the true input values 
finds $\Delta$FPP = 0.017 n - 0.067.

\begin{figure}[t]
\begin{centering}
\includegraphics[angle=0,totalheight=3.5in]{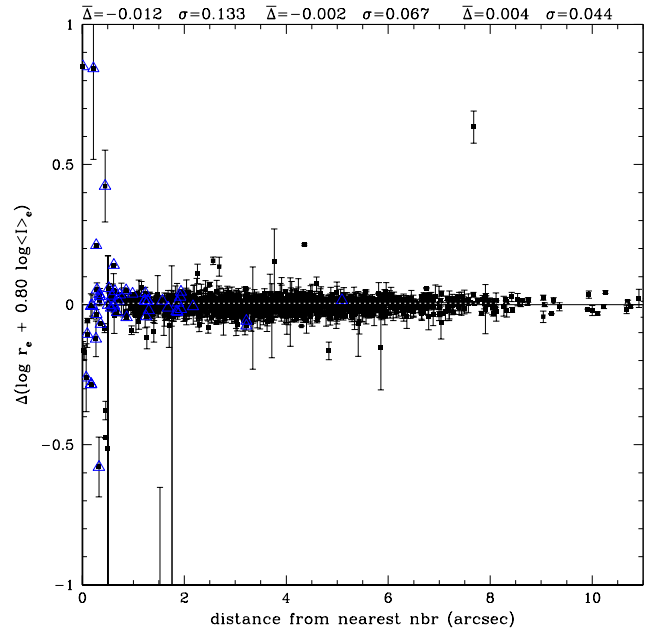}
\caption[Error in fundamental plane parameter]{Differences in r$^{1/4}$
law profile recovered parameter values for n $> 2$ galaxies as a
function of nearest neighbor distance. Points enclosed by triangles
have neighbors within 1 $r_e$. The average difference 
in recovered - input
values and standard deviation are provided for the ranges
d $< 2$, $2 <$ d $< 6$, and d $> 6$ arcsec.
\label{delnbr}}
\end{centering}
\end{figure}

We investigate other factors such as near neighbors which might affect the fits.
Simulated galaxies with close neighbors within 1 r$_e$ or within 1 arcsec 
display a greater deviation 
in the recovered vs. input values for the FPP than those galaxies with a 
nearest neighbor at a distance greater
than 2 arcsec.  In Figure \ref{delnbr}, we display the errors as a function
of nearest neighbor distance. Points enclosed by triangles have
neighbors within 1 r$_e$.  From the plot it is apparent that
these points have a larger dispersion than those with more 
distant neighbors.  The dispersion is 3 times as large for objects having
neighbors within 2 arcsec as compared to those with nearest neighbors
$> 6$ arcsec. 
We also test whether errors change as a function of r$_e$, magnitude,
or surface brightness 
(Table \ref{fppresults}).
We do find that random errors in the FPP may increase slightly at fainter magnitudes
and smaller sizes.
We find no trend as a function of the input c parameter, a measure of the galaxy core's 
intrinsic diskiness/boxiness.

Finally, we investigate how single profile fits are affected by multiple
component, bulge + disk, galaxies.  Forcing a single r$^{1/4}$ law fit
to a bulge embedded in a disk recovers a magnitude comparable to the sum of the two
components and a half-light radius that is larger than for the bulge alone.
When combined to form the FPP parameter, the average systematic 
offset from the FPP for the bulge component alone is $0.07$ 
with $\sigma = 0.09$ (Table \ref{devresults}).
This offset is larger than for the case of single profile 
galaxies although similar to the random error in single profile cases.
Obviously the extended exponential 
disk will affect the fit, but the resultant FPP is
similar to that of the bulge FPP alone within the expected uncertainties. 
Thus, we expect a r$^{1/4}$ law fit to a distant galaxy with large bulge and low 
surface brightness disk will 
recover primarily the bulge component with only a slight offset in the FPP. Likewise,
we would expect the spectroscopic measurement of the velocity dispersion to come
largely from light within the bulge.  For cases where faint disks are not obvious, we
expect FP measurements to primarily represent the bulge components. 
We conclude from our simulated galaxy analysis that while r$^{1/4}$ law fits may lead 
to systematic errors in 
individual parameters for real galaxies, we do not expect this to be a problem
for our FP analysis. We explore this further in Section \ref{ext}.

\subsubsection{Comparing real galaxy measurements}

\begin{figure}[t]
\begin{centering}
\plottwo{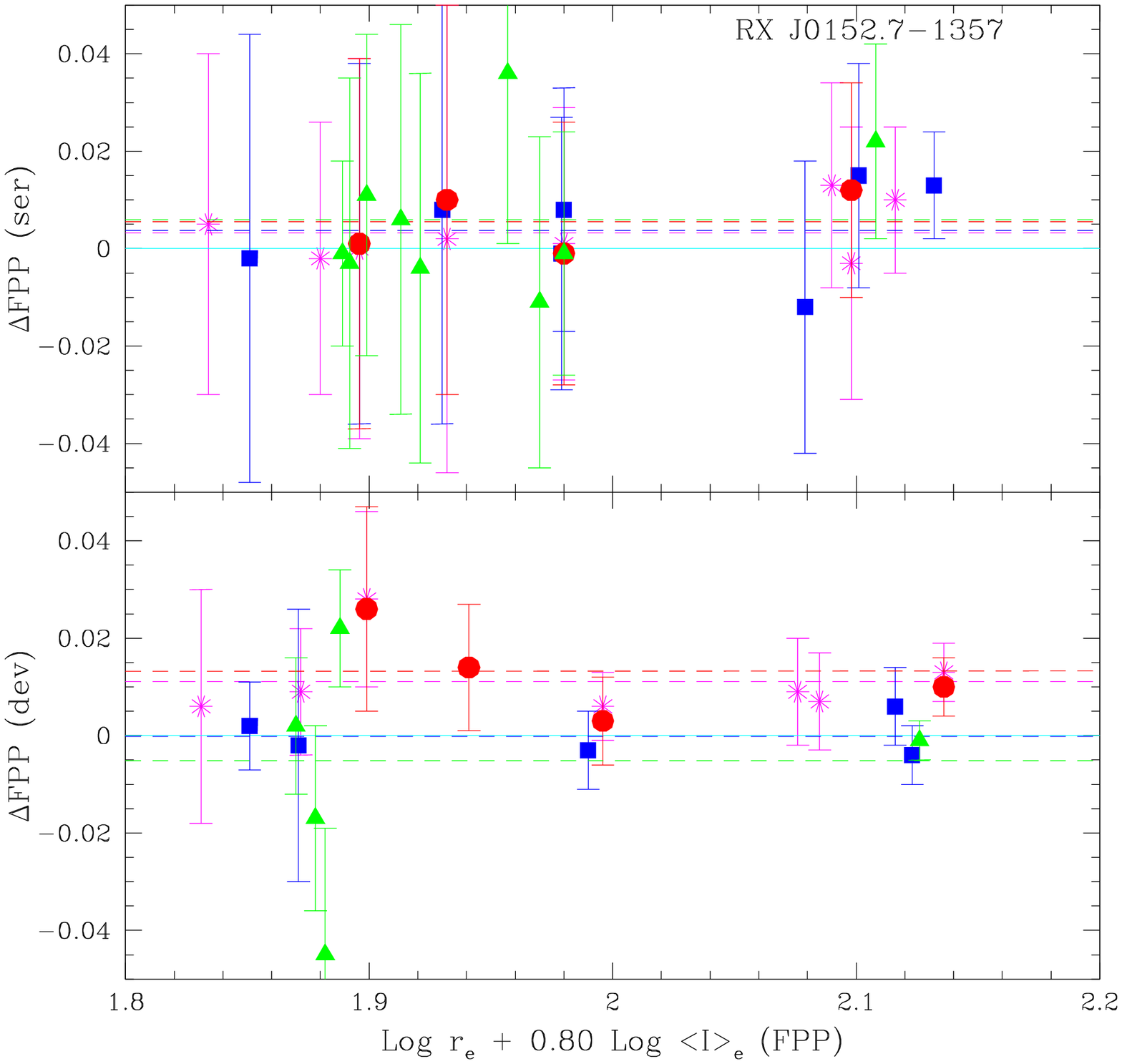}{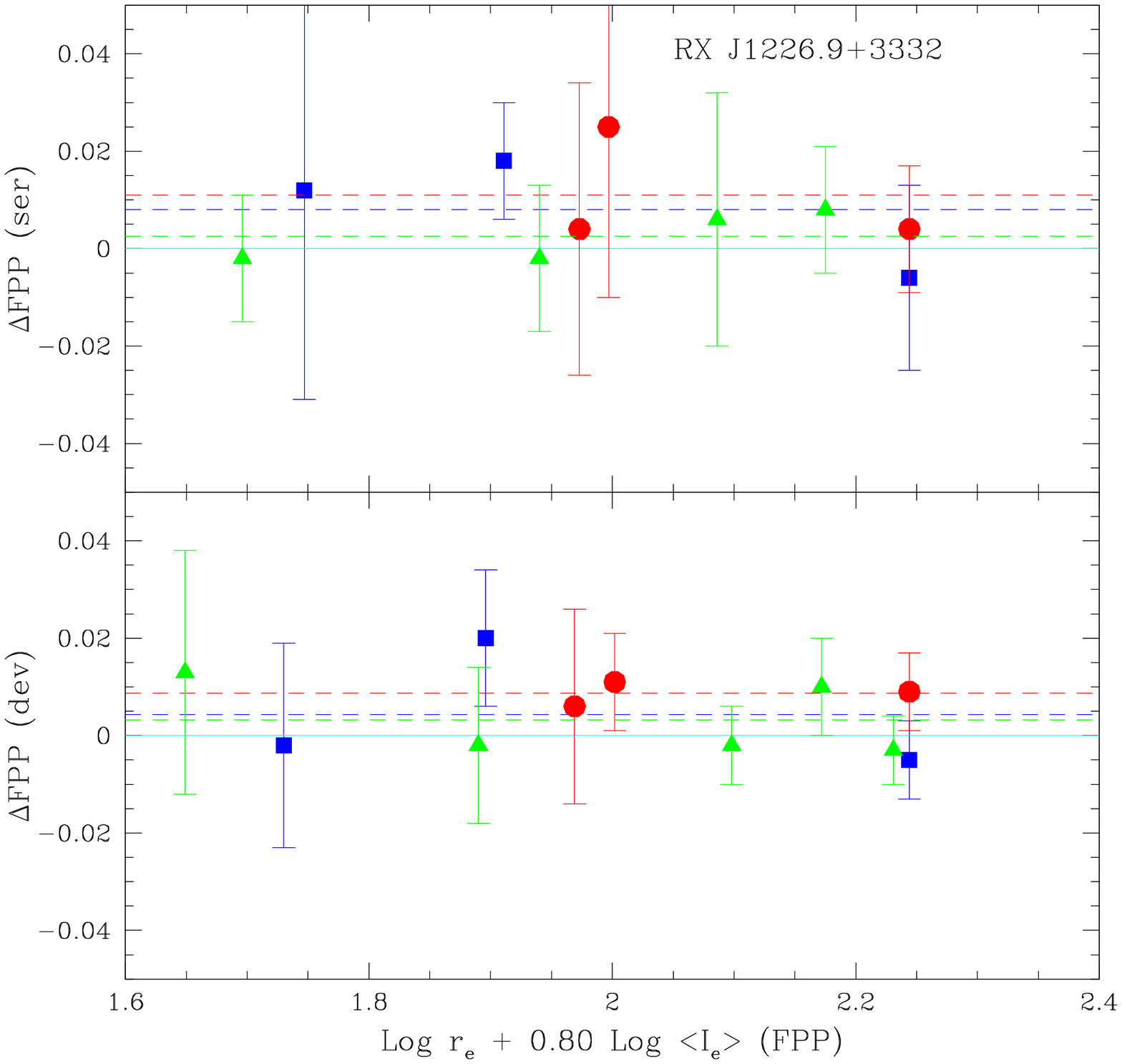}
\caption[Internal comparison of the FPP]{Comparison of the derived Fundamental
Plane parameter for galaxies observed in multiple images.   Plotted are
the differences in repeat measurements from overlapping pairs of images for both 
RX J0152.7-1367 and RX J1226.9+3332.  Average differences for each pair 
are shown by the dashed lines. 
\label{intcomp}}
\end{centering}
\end{figure}

For both RX J0152.7-1367 and RX J1226.9+3332, a number of galaxies were
observed multiple times in overlap regions from different telescope pointings.  As one
final test of our internal consistency we compare the derived 
structural and photometric properties from these multiple measurements.
In Figure \ref{intcomp}, we compare the difference in
derived FPP
for both Sersic and r$^{1/4}$ law profile fits in pairs of images.  Different
symbols denote different image pairs and the corresponding dashed lines
show the average difference for each pair.  
Average differences for pairs range from (absolute value) 0.0007 to
0.015 with standard deviations of 0.002 to 0.015.
These differences are generally smaller than GALFIT measurement uncertainties, and 
the scatter in these real measurement differences is smaller than the random
uncertainties
determined from simulated galaxies.  We do not find any worrisome systematic trends 
between images; all differences between images are within 
expected measurement uncertainties.

\subsection{External consistency}\label{ext}

\citet{blake} have used the same ACS dataset as used in this paper 
to study the structural
parameters and colors of 149 galaxies in RX J0152.7-1357.
They also perform surface brightness profile fitting using GALFIT
with a procedure very similar to ours.  Images are processed in
a similar manner, and like us, they fit all neighbors with
$i_{F775W} < 25 AB$ while masking objects fainter than this.
Differences in methodology are few but three significant
differences exist.  While we use a $9^{\prime\prime}$ Tiny Tim model PSF drizzled in the
same manner as our data, they choose to use empirical PSFs from
archival HST images.  They also impose an upper limit for the fitted
Sersic parameter, n, constraining the value to n $\leq 4$, whereas we
applied no constraints.  Because of the strong coupling between
Sersic function parameters, this will cause differences in the
measured values of both r$_{e}$ and total magnitude for galaxies
we find with n $> 4$.  Finally, while we hold the sky value constant
during the fitting at the SExtractor measured local value,
they simultaneously fit for the sky value with GALFIT, except in
cases where this led to poor fits.
                                                                                
\begin{figure}[t]
\begin{centering}
\includegraphics[angle=0,totalheight=4.5in]{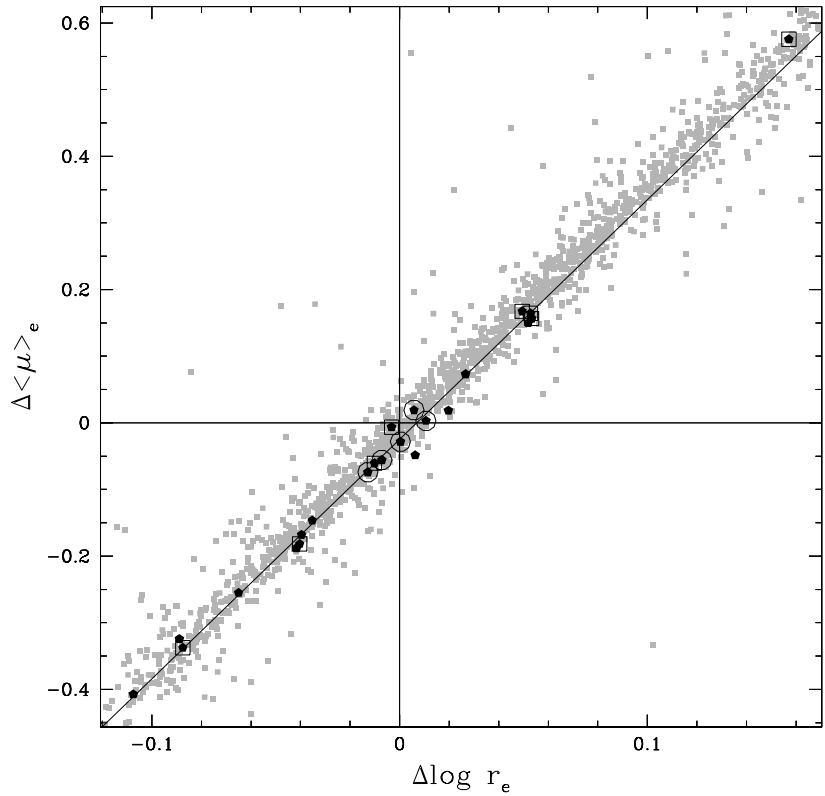}
\caption[Offset from Blakeslee ]{Offset in magnitude calibration between this work and
\citet{blake} (black). 
Galaxies with fitted Sersic n values differing by less than 0.1 are circled.
Boxed points correspond to galaxies found by us to have best fit n $> 4.2$.  The
best straight line fit to these points is shown. 
In gray, differences between our simulated galaxy input and
r$^{1/4}$ law recovered parameters.
\label{offsetbk}}
\end{centering}
\end{figure}

\begin{figure}[t]
\begin{centering}
\includegraphics[angle=0,totalheight=4.5in]{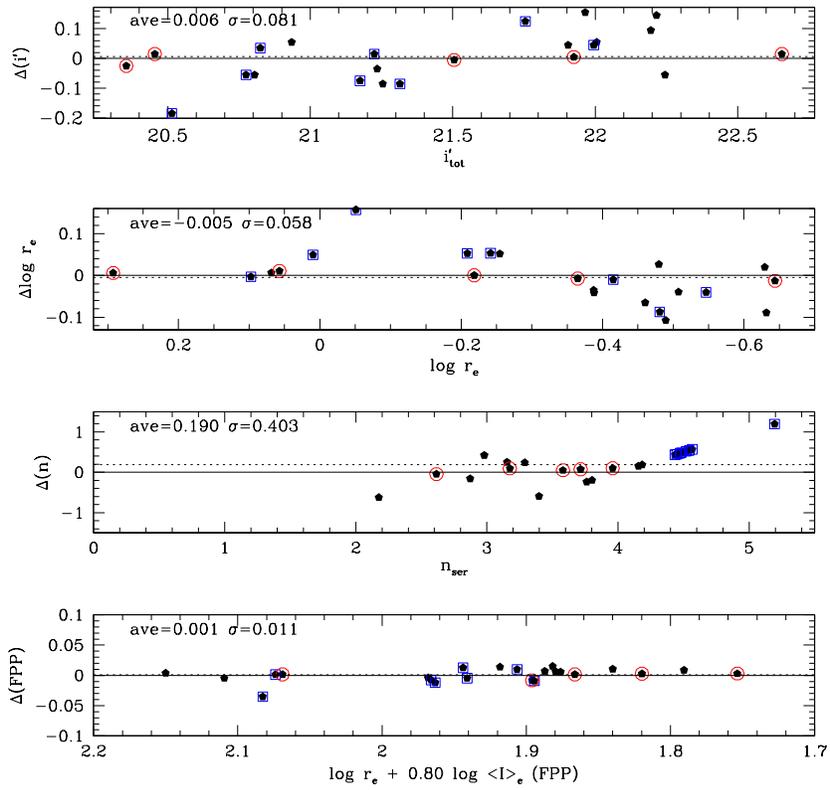}
\caption[Offset from Blakeslee ]{Differences between this work and
\citet{blake} in measured structural parameter values (in all cases (this work - Blakeslee))
after correction for magnitude zero point differences.  
Points corresponding
to galaxies best fitted with n $> 4.2$ are boxed.  Galaxies with measured Sersic
n differing by less than 0.1 are circled.
\label{bkcomp}}
\end{centering}
\end{figure}

We find a total of 26 objects in common, 3 of which have large discrepancies
in magnitude. These were galaxies that were not well fit with 
single Sersic functions and which had late type features evident in the 
residual images.  Excluding these,  we compare the measured 
Sersic surface brightness profile parameters of the remaining 23 galaxies
to the values listed in the Blakeslee et al. catalog.
We plot the differences in our measurements in a 
$\Delta \log r_e$ vs. $\Delta \langle \mu\rangle_e$ plane
(Figure \ref{offsetbk}).  From a linear fit to these data points we measure a magnitude 
zeropoint shift of $-0.025$ 
between the two works from the offset in the intercept from zero.  This is
in the sense that we find objects to be on average 0.025 mag brighter.
We find that large differences in measured surface brightness or size are correlated with 
large discrepancies in the recovered Sersic n parameter (see also Figure \ref{deln}).
Galaxies with measured n differing by
less than 0.1 for the two groups are circled.  
Correcting for the zeropoint shift we compare the difference in magnitudes, 
r$_{e}$, n, and the FPP.
Results are listed in Table \ref{bkresults} and displayed in Figure \ref{bkcomp}.

The increasing difference in measured n at n $>4$ is easily explained by
the constraints Blakeslee et al. impose on the Sersic n.
Blakeslee et al. chose not to use Tiny Tim models finding that 
analytical representations for the PSF overestimated
fitted r$_{e}$ values.  However, with our $9^{\prime\prime}$ model PSF we 
find r$_{e}$ values 
2\% smaller on average than their measurements.
Average differences in total measured magnitude and 
effective radius are insignificantly small compared to the rms scatter.
However, there appears to be a trend as a function of size
and magnitude.  We find brighter galaxies to be brighter than found by Blakeslee et al.
and fainter galaxies to be fainter.  Likewise, we measure larger sizes for larger
galaxies and smaller sizes for smaller galaxies in comparison to 
Blakeslee et al.  This can be attributed in part to different values recovered for the 
Sersic n parameter.  In five cases where we find similar values within 0.1 for measured n,
magnitude and size discrepancies are negligible regardless of size or magnitude.
For 10 objects measured by Blakeslee et al. to have $n = 4.0$ due to their constraint on
the Sersic index, and for which we find best
Sersic function fits with $n > 4.0$,  we compare the Blakeslee et al. results to our
r$^{1/4}$ law ($n = 4.0$) fits.  We find the average offsets in magnitude and size are similar  
but the random scatter is nearly half that of the differences in Sersic measurements
for these same galaxies.
For large galaxies, in cases where we find $n > 4.0$, the covariance of Sersic
function parameters generates fits with larger $r_e$ than found by Blakeslee et al. 
For small galaxies which are more strongly affected by the choice of
PSF, we speculate that our broad $9^{\prime\prime}$ PSF may
be the cause of the smaller size measurements as compared to Blakeslee et al. 
Due to the coupling of Sersic function parameters, a smaller recovered size results in a 
fainter recovered magnitude.    
In Figure \ref{offsetbk} we also show the parameter measurement errors in 
the $\Delta \langle \mu \rangle_e - \Delta \log r_e$ plane for our z $= 0.83$ 
simulations fitted 
with an r$^{1/4}$ law profile.  We find
the same coupling of errors and an offset in magnitude of 0.018, similar to that
found for the Blakeslee et al. comparison.  Since we used PSFs generated from real
stars to create the simulated galaxies and the $9^{\prime\prime}$ model PSF to fit the galaxies,
this further suggests that the broad PSF may be cause of the slight magnitude differences.

Because of the correlation between Sersic profile parameters, the FPP is insensitive 
to these small variations in size measurement.  When taking into account the small
magnitude zero point differences, we find the differences between
the Blakeslee et al. and our FPP measurement 
to be completely negligible and smaller than the systematic error of 0.006
determined from galaxy simulations.  The rms scatter in the differences
between our measurements is a factor of 5 smaller than the uncertainty 
expected from the simulations.  The systematic difference in the FPP measurement
of 0.009 is larger when not including the zero point shift, but is still much smaller
than our expected random uncertainty of 0.05 determined from simulations.

\subsection{Implications for FP measurements}\label{discFP}

Although we have fit each galaxy with both Sersic and r$^{1/4}$ law functions,
we use measurements of r$_{e}$ and surface brightness from the r$^{1/4}$ law 
fits to derive the FP since the galaxies 
in our low redshift comparison sample were originally fitted only with r$^{1/4}$ law 
profiles. Not all of our sample consists of pure r$^{1/4}$ law galaxies,
and the Sersic function fits find a wide range of values for the Sersic parameter.
However, only galaxies with n $> 2$ are used in our FP analysis
\citep{jorg06,jorg07,barr06}.
In this section we explore how galaxy parameter measurement uncertainties translate to 
FP measurement uncertainties and whether any systematic biases exist
in our methodology that would affect the derivation of the FP.

\begin{figure}[t]
\begin{centering}
\includegraphics[angle=0,totalheight=3.5in]{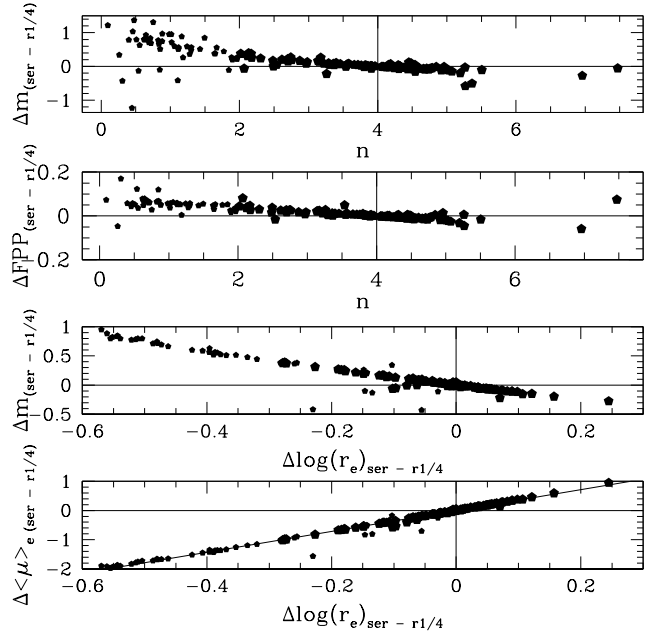}
\caption[Sersic vs R$^{1/4}$ law fits]{Top two panels:
differences in measured magnitude and FPP from Sersic
and r$^{1/4}$ law fits as a function of Sersic
n for all real sample galaxies in RX J0152.7-1357 and RX J1226.9+3332.
Next two panels: the difference in measured
magnitude and surface brightness vs. measured r$_e$ between Sersic
and r$^{1/4}$ law fits.  Objects best fit with Sersic n $> 2.0$ are
denoted by large symbols. We show a best straight line fit with slope 3.6 to the 
points in the bottom panel.
\label{mre}}
\end{centering}
\end{figure}

A plot of magnitude and FPP differences between GALFIT
Sersic and r$^{1/4}$ law fits of real galaxies vs Sersic index, n,
shows very good agreement for n $\sim4$ galaxies as expected (Figure \ref{mre}).
Many galaxies were best fit with n $< 4$ and, for these, the r$^{1/4}$ law measured magnitude
was as much as 1 mag brighter than that derived from the Sersic fits.
Although output parameters from Sersic and r$^{1/4}$ law fits often differed,
a plot of the difference in magnitude and in surface brightness vs. difference in log r$_e$ for
the two fits shows the parameters to be strongly correlated and
explains why differences in the FPP, although also correlated with Sersic n, 
are smaller. 
Galaxies with best Sersic fit $n > 2.0$ are displayed as larger points, but note that even
galaxies with low Sersic index follow the same linear relation.
Indeed, the FPP is reasonably robust against the form of the profile used to fit the
galaxies.  This effect has been noted by \citet{kidf00}, \citet{tgc01}, and \citet{l97}.
While the uncertainties in measured structural and photometric parameters may seem large
in cases where $n < 4.0$ according to best Sersic function fits, the FPP has 
a much lower $\sim0.01$ systematic error with $0.1$ random uncertainty.
We find that even when we fit a pure exponential
disk ($n=1$) with an r$^{1/4}$ law profile, the errors in the magnitude
and r$_{e}$ are large but the average error in the FPP is only 0.05.

\begin{figure}[t]
\begin{centering}
\includegraphics[angle=270,totalheight=3.5in]{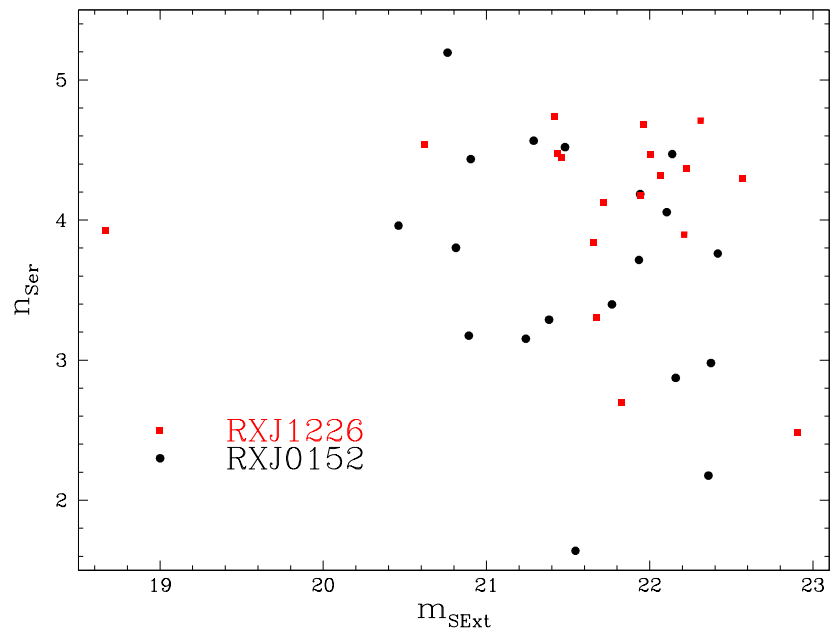}
\caption[Sersic n vs SExt mag]{Sersic index vs
SExtractor measured total magnitude for galaxies in our RX J0152.7-1357
and RX J1226.9+3332 FP samples \citep{jorg06,jorg07,barr06}.
\label{nmtrend}}
\end{centering}
\end{figure}

In our study of the FP we investigate whether the slope changes as a function
of redshift.  A change in slope would imply a change in M/L ratio as a function of
galaxy mass.  It is therefore critical to ensure our measurements are
not biased as a function of magnitude or size.  From our 
galaxy simulations we find some evidence for a slight increase of up to a factor
of 3 in the scatter of the FPP at smaller sizes ($< 1$ arcsec) and at fainter
magnitudes ($i > 22.5$).  However, this effect is small. More significantly,
we find a strong correlation between the uncertainty in the FPP and the
difference in measured vs. intrinsic Sersic index n.  Exponential disk 
galaxies with $n < 2$ fitted with r$^{1/4}$ law profiles have systematic 
offsets as large as the random uncertainty.
If there is a trend in real galaxy magnitudes with index n, this could 
affect the FP slope measurement.  In Figure \ref{nmtrend}, we display the
measured Sersic index n versus SExtractor total magnitude for the galaxies
in our RX J0152.7-1367 and RX J1226.9+3332 FP samples.  For the most part, the points 
exhibit random scatter. There is a hint of a trend in that the faintest
galaxies have a slightly lower average Sersic index n.  The correlation
coefficients for these two samples are small, only -0.31 and -0.14 
for RX J0152.7-1367 and RX J1226.9+3332 respectively. Although this slight 
correlation
coupled with the trend of increasing FPP measurement offset with decreasing n may
affect the FP slope, this effect is much smaller than 
the random uncertainties in the FPP.

It is also important to ensure our measurement uncertainties are insensitive
to redshift.  A redshift bias could be caused by, for example,
GALFIT parameter measurement uncertainties being systematically
different for objects with larger sizes or brighter magnitudes
at low redshift as compared with smaller, fainter objects at high redshift.
We do not find any evidence for this in our recovery of simulated galaxy parameters
in simulations of the high (z = 0.83) and low (z = 0.28) redshift
clusters.  The average magnitude and effective radius measurement uncertainties are
slightly larger for the low redshift sample
but scatter in the recovered values is about the same
for both clusters.   The
FPP has an average offset of recovered values from input which is insignificant
for both populations, and a random uncertainty which is larger for the lower redshift cluster
due to a small number of outliers with particularly poor fits.  
In our real galaxy sample
all failed fits are flagged and refitted or left out of the FP measurement.

Through this simulated galaxy analysis we have found a number of other factors 
which affect the galaxy structural parameter measurements.
Near neighbors affect profile fits and
based on simulations, we find that the highest surface brightness, most compact 
galaxies are affected the least. However, the only effect this has on lower surface
brightness galaxies is to increase the random uncertainty by up to factor
of 4 for only those few galaxies
which have a neighbor within 1 arcsec or 1 r$_{e}$.  
No systematic variation as a function of surface brightness is expected. 

Although we cannot simulate galaxies with a full range of substructures, nor
do we know the true distribution of such features in our real data, we
estimate uncertainties for galaxies with more complex structures with bulge + disk
simulations.  We find a systematic offset of $0.07$, much larger than the 0.01
for simulated galaxies generated with a single profile, with a slightly larger rms 
scatter of 0.10.  
With a large sample of galaxies exhibiting complex morphologies, we would expect 
measurements of the FP to be affected by the larger uncertainties. 
However, we check
the goodness of all fits from residual images.  Any features not
modeled by the simple Sersic/r$^{1/4}$ law profile fits will show up
in these images as imperfect subtraction from the galaxy model. For most
of our galaxies, subtraction is good without any obvious sign
of extra arms, twisted isophotes, low surface brightness disks, or any
other structures.  Galaxies with such features tend to be best fit with Sersic 
$n < 2.0$, and these objects are not included in our FP measurements.

In conclusion,
we take final uncertainties in the FPP to be $\sim 0.1$ based on our
simulations, dominated by random uncertainties.
We find no systematic offsets between recovered and input values for the
FPP as a function of magnitude or half-light radius from our galaxy simulations.
While we do find uncertainties are affected by near neighbors or
Sersic index n, we do not expect the shape or tilt of the FP to be
affected by these uncertainties, unless for example all lower mass galaxies have exponential
profiles, which we see no evidence for.   We find that our measurement uncertainties should
not vary with redshift unless, for example, higher redshift galaxies contain more
late-type morphological features.  Although this is a possibility, we would expect to
spot this in our residual fitting images.  
We therefore do
not find any evidence for measurement uncertainties or systematic
biases introduced by our data reduction methods to affect our derived FP as a function
of redshift, galaxy size, or magnitude.  We expect the measured tilts
to be intrinsic values for each cluster.

\section{Summary}\label{sum}

Using new and archival HST/ACS high resolution images observed in
the $F775W$ or $F814W$ bands, we have
measured structural and photometric properties for galaxies
in the clusters RX J0142.0+2131, RX J0152.7-1367, and RX J1226.9+3332.
These data are being used to investigate the Fundamental Plane as a 
function of redshift, mass, and environment as part of our 
Gemini/HST Galaxy Cluster Project aimed at understanding cluster galaxy
evolution.  In this work we 
have described our data processing and analysis methods, presented catalogs
of parameters measured from 2-D surface brightness profile fitting,
and discussed expected uncertainties in our measurements and in the quantities
which go into the measurement of the FP.

We determine average uncertainties in measured parameters for n $> 2$ galaxies 
from galaxy simulation tests.  We define systematic errors and random uncertainties 
as the mean and rms scatter in the difference between recovered and input 
simulation values.
For measured total magnitudes, these are  $-0.1\pm0.3$ for r$^{1/4}$ law profiles.
The average error in size, $\log$ r$_{e}$, is $0.1\pm0.2$.   Random uncertainties 
are larger than any systematic offsets from intrinsic values.
Sersic profile fits do slightly better than r$^{1/4}$ law profiles
in recovering the input values of the
simulated galaxies with magnitude errors of $\sim0.03\pm0.3$
and errors in $\log$ r$_{e}$ of $\sim -0.02\pm0.2$.  However, the rms scatter
is similar for fits with either profile.
                                                                                                    
Due to the combination of parameters 
that enter the FP, $\log$ r$_{e}$ + $\beta \log \langle I\rangle_{e}$,
we find structural and photometric uncertainties
to have little effect on the FP measurement.
We take final uncertainties in this FP parameter to 
be $\sim 0.1$, dominated by random error.
We find no systematic offsets between recovered and input values for the
FPP in our galaxy simulations as a function of magnitude or effective
radius.  While we do find uncertainties are affected by near neighbors or
Sersic index n, we do not expect the shape or tilt of the FP to be
affected by these uncertainties, unless for example all lower mass galaxies 
have exponential
profiles, something not seen in our samples.   
We find that our measurement uncertainties should
be invariant with redshift unless, for example, higher redshift galaxies contain more
late-type morphological features.  While this is a possibility, we would expect to
spot this in our residual fitting images, so do not expect this to significantly
affect our results.
We conclude that the methods used to derive the effective
parameters r$_e$ and $\langle \mu \rangle_e$ do not introduce any significant
bias in our FP measurements,  or in the results
presented for these three clusters in \citet{jorg06,jorg07,barr06}.

\acknowledgments

The authors would like to express their gratitude to the STScI helpdesk
for their quick responses to questions about, and patches
for, MULTIDRIZZLE, Chien Peng for
assistance with GALFIT, Kathleen Labrie for PyRAF installation help,
Alexander Fritz for helpful comments on the manuscript,
and the anonymous referee for many helpful suggestions for improving this paper.
This work is based on observations made with the NASA/ESA Hubble
Space Telescope. K. C., I. J., and K. F. acknowledge support from grant
HST-GO-09770.01 from STScI. STScI is operated by AURA, Inc., under
NASA contract NAS 5-26555.  Supported by the Gemini Observatory, which is
operated by the Association of Universities for Research in Astronomy, Inc.,
on behalf of the international Gemini partnership of Argentina,
Australia, Brazil, Canada, Chile, the United Kingdom, and the United States of America.

\bibliographystyle{apj}
\bibliography{kc}


\clearpage
\begin{deluxetable}{lllll}
\tabletypesize{\normalsize}
\tablewidth{0pt}
\tablecaption{HST/ACS Imaging Data\label{imagetab}}
\tablehead{
\colhead{Cluster} &
\colhead{\# fields} &
\colhead{filter} &
\colhead{total t$_{exp}$(s)} &
\colhead{Program ID} \\
}
\startdata
 RXJ0142.0+2131 & 2 & F775W ($i'$) &  4420 & 9770 \\
RXJ0152.7+1357  &  4 & F625W ($r'$) & 4750 & 9290  \\
  &  4 & F775W ($i'$) & 4800 &  \\
  &  4 & F850LP ($z'$) & 4750  &   \\
RXJ1226.9+3332  &  4 & F606W ($R$) & 4000 & 9033 \\
  &  4 & F814W ($I$) & 4000 &  \\
\enddata
\end{deluxetable}

\clearpage

\clearpage
\begin{deluxetable}{lrrrr}
\tabletypesize{\normalsize}
\tablewidth{0pt}
\tablecaption{GALFIT Median Measurement Uncertainties\label{mederr}}
\tablehead{
\colhead{} &
\colhead{z$=0.83-0.89$} &
\colhead{} &
\colhead{z$=0.28$} &
\colhead{} \\
\colhead{} &
\colhead{Sersic fit} &
\colhead{r$^{1/4}$} &
\colhead{Sersic} &
\colhead{r$^{1/4}$} \\
}
\startdata
Magnitude & 0.010 & 0.006 & 0.003 & 0.001 \\
r$_e$  &  0.009 & 0.009  & 0.009 & 0.008  \\
$\langle\mu \rangle_e$  &  0.040 & 0.025 & 0.013 & 0.008 \\
\enddata

\tablecomments{Uncertainties are similar for the two high redshift clusters
and are therefore combined in this table.}
\end{deluxetable}

\clearpage

\clearpage
\begin{deluxetable}{llllrrrrrrrrrl}
\rotate
\tabletypesize{\scriptsize}
\tablewidth{0pt}
\tablecaption{RXJ0152.7-1357: Photometric and Structural Parameters\label{rxj0152cat}}
\tablehead{
\colhead{ID} &
\colhead{RA} &
\colhead{Dec} &
\colhead{N$_{meas}$} &
\colhead{m$_{tot,dev}$} &
\colhead{$\log(r_{e})_{dev}$} &
\colhead{$\langle\mu \rangle_{e,dev}$} &
\colhead{m$_{tot,ser}$} &
\colhead{$\log(r_{e})_{ser}$} &
\colhead{$\langle\mu \rangle_{e,ser}$} &
\colhead{n$_{ser}$} &
\colhead{PA$_{ave}$} &
\colhead{$\epsilon_{ave}$} &
\colhead{Member\tablenotemark{a}} \\
\colhead{} & 
\colhead{(2000)} &
\colhead{} &
\colhead{} &
\colhead{(F775W)} &
\colhead{(arcsec)} &
\colhead{mag arcsec$^{-2}$} &
\colhead{(F775W)} &
\colhead{(arcsec)} &
\colhead{mag arcsec$^{-2}$} &
\colhead{} &
\colhead{} &
\colhead{} &
\colhead{} \\
}
\startdata

193\tablenotemark{\dag} &1:52:50.80 &-13:55:28.9 & 2 & 20.25 & 0.115 & 22.82 & 20.64 & -0.161 & 21.83 & 2.13 & 34.3 & 0.07 & 0 \\
264 &1:52:44.66 &-13:55:37.3 & 1 & 21.17 & -0.181 & 22.26 & 21.17 & -0.167 & 22.89 & 4.33 & -29.2 & 0.47 & 0 \\
338 &1:52:43.33 &-13:55:44.4 & 1 & 21.93 & -0.447 & 21.69 & 22.19 & -0.632 & 21.02 & 2.18 & -17.4 & 0.33 & 1 \\
346 &1:52:37.42 &-13:55:50.1 & 2 & 21.11 & -0.408 & 21.06 & 21.21 & -0.480 & 20.80 & 3.29 & 133.8 & 0.25 & 1 \\
422 &1:52:34.59 &-13:55:58.8 & 1 & 21.80 & -0.494 & 21.32 & 21.79 & -0.487 & 21.34 & 4.06 & 73.7 & 0.09 & 1 \\
460 &1:52:36.11 &-13:56:08.5 & 1 & 20.87 & -0.365 & 21.04 & 20.91 & -0.387 & 20.96 & 4.16 & 101.9 & 0.25 & 0 \\
523 &1:52:42.38 &-13:56:18.7 & 2 & 21.10 & -0.360 & 21.30 & 21.20 & -0.415 & 21.12 & 4.48 & -10.2 & 0.14 & 1 \\
566 &1:52:38.03 &-13:56:28.1 & 2 & 20.90 & -0.097 & 22.41 & 20.75 & 0.010 & 22.79 & 4.54 & 1.5 & 0.53 & 1 \\
627 &1:52:38.48 &-13:56:33.6 & 2 & 21.81 & -0.350 & 22.05 & 21.88 & -0.388 & 21.93 & 4.18 & -2.3 & 0.25 & 1 \\
643 &1:52:45.60 &-13:56:40.0 & 2 & 21.41 & -0.168 & 22.58 & 21.48 & -0.218 & 22.38 & 3.58 & 132.5 & 0.46 & 1 \\
737 &1:52:45.77 &-13:56:46.1 & 2 & 21.85 & -0.329 & 22.20 & 21.90 & -0.365 & 22.06 & 3.71 & 54.2 & 0.08 & 1 \\
766 &1:52:45.83 &-13:56:59.2 & 2 & 20.33 & 0.063 & 22.63 & 20.33 & 0.057 & 22.61 & 3.96 & 32.6 & 0.22 & 1 \\
776 &1:52:38.48 &-13:56:52.5 & 1 & 21.44 & -0.816 & 19.35 & 21.45 & -0.823 & 19.33 & 3.86 & 54.8 & 0.61 & 1 \\
813 &1:52:44.97 &-13:57:04.2 & 4 & 20.68 & -0.209 & 21.63 & 20.49 & -0.051 & 22.22 & 5.20 & 0.5 & 0.64 & 1 \\
896\tablenotemark{\dag} &1:52:36.99 &-13:57:10.1 & 1 & 22.29 & -0.179 & 23.39 & 21.87 & -0.234 & 22.68 & 0.31 & -13.3 & 0.32 & 1 \\
908 &1:52:43.74 &-13:57:19.4 & 4 & 20.80 & 0.074 & 23.16 & 20.78 & 0.067 & 23.11 & 3.80 & 50.9 & 0.21 & 1 \\
1027 &1:52:43.32 &-13:57:26.7 & 4 & 21.83 & -0.413 & 21.76 & 21.94 & -0.489 & 21.49 & 3.40 & 13.3 & 0.24 & 1 \\
1085 &1:52:42.94 &-13:57:35.0 & 4 & 21.23 & -0.273 & 21.86 & 21.15 & -0.208 & 22.10 & 4.52 & 72.2 & 0.27 & 1 \\
1110 &1:52:39.93 &-13:57:42.6 & 2 & 21.09 & -0.143 & 22.37 & 21.23 & -0.254 & 21.95 & 3.15 & -26.2 & 0.26 & 1 \\
1159\tablenotemark{\dag} &1:52:36.18 &-13:57:48.8 & 1 & 21.61 & -0.411 & 21.55 & 21.73 & -0.481 & 21.32 & 4.45 & -7.9 & 0.47 & 1 \\
1210 &1:52:42.83 &-13:57:55.3 & 2 & 21.81 & -0.390 & 21.86 & 21.98 & -0.508 & 21.43 & 2.87 & 122.5 & 0.29 & 1 \\
1245 &1:52:43.57 &-13:58:00.0 & 1 & 21.18 & 0.064 & 23.50 & 21.93 & -0.416 & 21.84 & 0.88 & 21.6 & 0.27 & 0 \\
1299\tablenotemark{\dag} &1:52:47.34 &-13:59:26.1 & 1 & 21.28 & -0.071 & 22.92 & 21.28 & -0.134 & 22.60 & 2.49 & -22.3 & 0.48 & 1 \\
1385\tablenotemark{\dag} &1:52:39.36 &-13:59:04.5 & 1 & 22.10 & 0.021 & 24.20 & 21.69 & -0.209 & 22.64 & 1.11 & 79.4 & 0.34 & 1 \\
1458 &1:52:39.64 &-13:58:56.6 & 2 & 21.94 & -0.541 & 21.22 & 21.97 & -0.547 & 21.23 & 4.47 & 21.2 & 0.29 & 1 \\
1494 &1:52:39.08 &-13:58:48.8 & 2 & 18.44 & -0.252 & 19.17 & 18.41 & -0.223 & 19.28 & 4.44 & 108.0 & 0.51 & 0 \\
1507 &1:52:34.48 &-13:58:42.2 & 1 & 22.14 & -0.436 & 21.94 & 22.17 & -0.460 & 21.85 & 3.76 & 61.6 & 0.56 & 1 \\
1567 &1:52:39.62 &-13:58:26.7 & 2 & 20.17 & 0.472 & 24.51 & 20.43 & 0.292 & 23.88 & 3.17 & 50.4 & 0.20 & 1 \\
1590 &1:52:38.87 &-13:58:32.0 & 2 & 22.09 & -0.531 & 21.42 & 22.22 & -0.629 & 21.06 & 2.98 & 85.7 & 0.34 & 1 \\
1614 &1:52:51.96 &-13:58:17.1 & 1 & 20.88 & 0.036 & 23.05 & 20.80 & 0.098 & 23.28 & 4.43 & 62.1 & 0.15 & 1 \\
1682 &1:52:51.96 &-13:58:15.6 & 1 & 21.16 & -0.163 & 22.34 & 21.54 & -0.446 & 21.30 & 1.64 & 127.7 & 0.49 & 1 \\
1811 &1:52:38.63 &-13:59:20.8 & 1 & 22.43 & -0.495 & 21.94 & 22.63 & -0.644 & 21.40 & 2.62 & 3.3 & 0.37 & 1 \\
1920\tablenotemark{\dag} &1:52:39.70 &-13:59:14.3 & 1 & 21.94 & -0.013 & 23.87 & 21.84 & -0.160 & 23.03 & 0.85 & -33.4 & 0.92 & 1 \\
1935 &1:52:41.88 &-13:59:53.6 & 1 & 21.36 & -0.295 & 21.87 & 21.29 & -0.242 & 22.07 & 4.57 & 10.4 & 0.18 & 1 \\
1970\tablenotemark{\dag} &1:52:48.03 &-13:59:58.6 & 1 & 19.78 & 0.389 & 23.72 & 20.58 & -0.166 & 21.74 & 0.62 & 57.4 & 0.78 & 0 \\
2042\tablenotemark{\dag} &1:52:42.38 &-13:59:46.6 & 1 & 18.12 & 0.958 & 24.90 & 19.43 & 0.212 & 22.48 & 0.75 & -40.3 & 0.17 & 0 \\

\enddata
                                                                                                                                                         
\tablenotetext{a}{1: galaxy is considered a member of RX J0152.7-1357; 0: galaxy is not a member}
\tablenotetext{\dag}{Galaxy exhibits late-type structure and is not well fit with
an r$^{1/4}$ law profile.}
                                                                                                                                                         
\end{deluxetable}

\clearpage

\clearpage
\begin{deluxetable}{llllrrrrrrrrrl}
\rotate
\tabletypesize{\scriptsize}
\tablewidth{0pt}
\tablecaption{RXJ1226.9+3332: Photometric and Structural Parameters\label{rxj1226cat}}
\tablehead{
\colhead{ID} &
\colhead{RA} &
\colhead{Dec} &
\colhead{N$_{meas}$} &
\colhead{m$_{tot,dev}$} &
\colhead{$\log(r_{e})_{dev}$} &
\colhead{$\langle\mu \rangle_{e,dev}$} &
\colhead{m$_{tot,ser}$} &
\colhead{$\log(r_{e})_{ser}$} &
\colhead{$\langle\mu \rangle_{e,ser}$} &
\colhead{n$_{ser}$} &
\colhead{PA$_{ave}$} &
\colhead{$\epsilon_{ave}$} &
\colhead{Member\tablenotemark{a}} \\
\colhead{} & 
\colhead{(2000)} &
\colhead{} &
\colhead{} &
\colhead{(F814W)} &
\colhead{(arcsec)} &
\colhead{mag arcsec$^{-2}$} &
\colhead{(F814W)} &
\colhead{(arcsec)} &
\colhead{mag arcsec$^{-2}$} &
\colhead{} &
\colhead{} &
\colhead{} &
\colhead{} \\
}
\startdata

18 & 12:27:06.81 & 33:35:29.73 & 1 & 21.50 & -0.615 & 20.42 & 21.54 & -0.649 & 20.29 & 3.55 & -84.2 & 0.41 & 0 \\
38 & 12:27:05.65 & 33:35:28.64 & 1 & 21.51 & -0.097 & 23.01 & 21.95 & -0.417 & 21.86 & 1.77 & 57.0 & 0.51 & 1 \\
55 & 12:26:53.16 & 33:35:13.69 & 1 & 21.41 & -0.523 & 20.79 & 21.44 & -0.541 & 20.73 & 4.45 & 5.1 & 0.18 & 1 \\
56\tablenotemark{\dag} & 12:27:08.86 & 33:35:19.88 & 1 & 21.37 & -0.222 & 22.25 & 21.59 & -0.384 & 21.66 & 2.70 & 8.8 & 0.35 & 1 \\
91\tablenotemark{\dag} & 12:27:06.64 & 33:35:08.25 & 1 & & & & 23.36 & -0.496 & 22.87 & 0.69 & -16.2 & 0.67 & 0 \\   
104 & 12:26:59.78 & 33:35:02.07 & 1 & 21.89 & -0.374 & 22.01 & 21.85 & -0.339 & 22.15 & 4.37 & 29.2 & 0.02 & 1 \\
122\tablenotemark{\dag} & 12:27:01.32 & 33:34:56.27 & 1 & 22.01 & -0.442 & 21.79 & 22.09 & -0.503 & 21.56 & 3.36 & 31.4 & 0.64 & 1 \\
132 & 12:26:47.77 & 33:34:48.31 & 1 & 21.24 & -0.518 & 20.64 & 21.17 & -0.461 & 20.85 & 4.66 & 42.0 & 0.64 & 0 \\
138 & 12:27:00.63 & 33:34:46.95 & 1 & 20.93 & -0.789 & 18.98 & 20.82 & -0.707 & 19.28 & 5.50 & -81.4 & 0.32 & 0 \\
154 & 12:26:47.80 & 33:34:44.64 & 1 & 22.51 & -1.014 & 19.42 & 22.38 & -0.952 & 19.61 & 8.06 & -19.0 & 0.24 & 0 \\
178\tablenotemark{\dag} & 12:26:48.35 & 33:34:40.22 & 1 & 21.33 & -0.155 & 22.55 & 21.90 & -0.542 & 21.18 & 1.12 & -90.2 & 0.71 & 0 \\
185 & 12:26:53.32 & 33:34:36.17 & 1 & 21.18 & -0.329 & 21.52 & 21.13 & -0.294 & 21.65 & 4.24 & 38.2 & 0.13 & 0 \\
203\tablenotemark{\dag} & 12:26:59.35 & 33:34:19.92 & 1 & 20.17 & 0.499 & 24.66 & 20.97 & -0.005 & 22.93 & 0.63 & 6.5 & 0.03 & 0 \\
220 & 12:26:55.16 & 33:34:25.53 & 2 & 21.58 & -0.466 & 21.24 & 21.51 & -0.410 & 21.45 & 4.52 & -28.1 & 0.21 & 0 \\
229 & 12:27:07.35 & 33:34:24.34 & 1 & 21.89 & -0.353 & 22.11 & 21.74 & -0.232 & 22.57 & 5.07 & 51.6 & 0.11 & 1 \\
245 & 12:27:07.42 & 33:34:19.73 & 1 & 22.00 & -0.609 & 20.94 & 21.95 & -0.575 & 21.07 & 4.43 & -27.3 & 0.32 & 0 \\
247\tablenotemark{\dag} & 12:26:51.34 & 33:34:14.92 & 1 & 18.94 & 0.337 & 22.62 & 19.60 & -0.125 & 20.97 & 0.83 & -48.5 & 0.40 & 0 \\
249\tablenotemark{\dag} & 12:26:52.08 & 33:34:12.47 & 1 & 19.81 & 0.202 & 22.81 & 20.44 & -0.194 & 21.46 & 0.56 & -4.6 & 0.67 & 0 \\
293 & 12:27:05.85 & 33:34:05.23 & 1 & 21.35 & -0.301 & 21.84 & 21.34 & -0.278 & 21.94 & 4.48 & 23.4 & 0.29 & 1 \\
295\tablenotemark{\dag} & 12:27:02.18 & 33:34:06.41 & 1 & 21.42 & -0.009 & 23.37 & 21.36 & -0.111 & 22.80 & 2.07 & -14.0 & 0.07 & 1 \\
309 & 12:27:04.45 & 33:34:06.81 & 1 & 22.62 & -0.944 & 19.89 & 22.63 & -0.948 & 19.88 & 3.92 & -45.0 & 0.53 & 1 \\
310 & 12:26:53.25 & 33:34:05.72 & 1 & 22.42 & -0.912 & 19.85 & 22.37 & -0.875 & 19.98 & 4.77 & 47.1 & 0.45 & 1 \\
316\tablenotemark{\dag} & 12:26:59.58 & 33:34:04.32 & 1 & 21.56 & 0.620 & 26.65 & 22.93 & -0.206 & 23.89 & 0.48 & -77.7 & 0.47 & 0 \\
329\tablenotemark{\dag} & 12:26:54.54 & 33:33:56.56 & 1 & 20.63 & 0.627 & 25.76 & 21.85 & -0.146 & 23.11 & 0.10 & -71.9 & 0.21 & 0 \\
333\tablenotemark{\dag} & 12:27:00.44 & 33:33:59.17 & 1 & 22.14 & 0.239 & 25.32 & 23.09 & -0.362 & 23.27 & 0.98 & 20.8 & 0.44 & 0 \\
347\tablenotemark{\dag} & 12:27:04.39 & 33:33:52.62 & 1 & 20.23 & -0.059 & 21.93 & 20.74 & -0.428 & 20.60 & 1.12 & -52.1 & 0.56 & 0 \\
349 & 12:27:01.33 & 33:33:53.42 & 1 & 20.78 & -0.571 & 19.91 & 20.98 & -0.713 & 19.40 & 1.90 & -29.0 & 0.25 & 0 \\
359\tablenotemark{\dag} & 12:26:54.38 & 33:33:51.04 & 1 & 20.82 & 0.187 & 23.74 & 21.53 & -0.300 & 22.02 & 0.69 & -77.0 & 0.67 & 0 \\
374 & 12:27:02.07 & 33:33:43.63 & 1 & 20.94 & -0.032 & 22.77 & 21.50 & -0.428 & 21.35 & 1.30 & -47.9 & 0.49 & 0 \\
386\tablenotemark{\dag} & 12:26:54.48 & 33:33:40.70 & 1 & 21.54 & -0.139 & 22.84 & 21.92 & -0.411 & 21.85 & 2.02 & 7.7 & 0.57 & 0 \\
408 & 12:26:54.25 & 33:33:34.78 & 1 & 22.82 & -1.217 & 18.72 & 22.24 & -0.836 & 20.05 & 5.26 & 16.6 & 0.38 & 0 \\
423 & 12:26:53.15 & 33:33:31.39 & 1 & 21.83 & -0.477 & 21.43 & 21.71 & -0.378 & 21.81 & 5.03 & -34.4 & 0.19 & 1 \\
441 & 12:26:53.86 & 33:33:28.92 & 1 & 22.60 & -0.698 & 21.10 & 22.77 & -0.816 & 20.67 & 2.48 & -8.0 & 0.59 & 1 \\
446 & 12:27:06.76 & 33:33:27.15 & 1 & 21.74 & -0.647 & 20.50 & 21.75 & -0.654 & 20.47 & 3.91 & -14.9 & 0.26 & 1 \\
452\tablenotemark{\dag} & 12:26:53.99 & 33:33:23.08 & 1 & 22.01 & -0.203 & 22.99 & 21.90 & -0.232 & 22.73 & 1.85 & -33.1 & 0.54 & 1 \\
462 & 12:26:52.47 & 33:33:24.18 & 1 & 22.45 & -0.565 & 21.62 & 22.35 & -0.475 & 21.96 & 5.01 & -60.4 & 0.06 & 1 \\
470 & 12:26:54.90 & 33:33:24.48 & 1 & 22.95 & -0.764 & 21.12 & 22.96 & -0.774 & 21.08 & 3.84 & 54.1 & 0.28 & 1 \\
491 & 12:26:57.55 & 33:33:14.05 & 2 & 21.36 & -0.300 & 21.85 & 21.38 & -0.319 & 21.78 & 3.84 & 24.4 & 0.21 & 1 \\
499\tablenotemark{\dag} & 12:27:07.72 & 33:33:14.32 & 2 & 22.05 & -0.127 & 23.40 & 22.64 & -0.535 & 21.96 & 1.24 & 16.2 & 0.61 & 0 \\
500\tablenotemark{\dag} & 12:26:53.55 & 33:33:10.10 & 1 & 20.71 & -0.065 & 22.37 & 21.24 & -0.443 & 21.01 & 1.12 & -38.4 & 0.68 & 0 \\
512 & 12:26:55.67 & 33:33:13.13 & 1 & 22.77 & -0.820 & 20.66 & 22.86 & -0.880 & 20.45 & 2.97 & -32.0 & 0.53 & 1 \\
523 & 12:26:57.18 & 33:33:05.06 & 1 & 21.56 & -0.182 & 22.64 & 21.50 & -0.145 & 22.77 & 3.91 & -5.9 & 0.26 & 0 \\
528\tablenotemark{\dag} & 12:27:04.49 & 33:33:07.77 & 2 & 20.91 & 0.071 & 23.25 & 21.42 & -0.283 & 22.00 & 1.43 & -80.4 & 0.79 & 1 \\
529 & 12:27:01.38 & 33:33:04.80 & 2 & 21.17 & -0.418 & 21.07 & 21.08 & -0.343 & 21.35 & 4.74 & -68.8 & 0.22 & 1 \\
534 & 12:26:56.83 & 33:33:06.30 & 1 & 21.98 & -0.590 & 21.02 & 22.06 & -0.654 & 20.78 & 3.22 & -72.3 & 0.53 & 1 \\
547\tablenotemark{\dag} & 12:26:54.23 & 33:32:53.61 & 1 & 21.12 & 0.395 & 25.09 & 22.01 & -0.165 & 23.17 & 1.16 & -23.8 & 0.42 & 1 \\
557 & 12:27:07.73 & 33:32:55.59 & 1 & 22.10 & -0.689 & 20.64 & 22.08 & -0.678 & 20.68 & 4.17 & 73.8 & 0.24 & 1 \\
563 & 12:26:58.29 & 33:32:49.00 & 3 & 18.70 & 0.662 & 24.00 & 18.71 & 0.653 & 23.96 & 3.92 & -84.6 & 0.38 & 1 \\
572 & 12:27:02.65 & 33:32:57.47 & 2 & 21.92 & -0.944 & 19.19 & 21.88 & -0.914 & 19.30 & 4.79 & 60.1 & 0.51 & 0 \\
593\tablenotemark{\dag} & 12:26:57.08 & 33:32:53.54 & 2 & 22.15 & -0.585 & 21.21 & 22.36 & -0.732 & 20.69 & 1.97 & 12.0 & 0.65 & 1 \\
602 & 12:26:53.01 & 33:32:49.89 & 1 & 21.73 & -0.892 & 19.25 & 21.69 & -0.883 & 19.27 & 5.26 & -51.9 & 0.35 & 1 \\
608 & 12:26:50.63 & 33:32:46.54 & 1 & 21.53 & -0.520 & 20.92 & 21.42 & -0.429 & 21.26 & 4.92 & -82.5 & 0.12 & 1 \\
630 & 12:26:57.35 & 33:32:46.85 & 2 & 22.88 & -0.732 & 21.22 & 22.94 & -0.762 & 21.12 & 3.29 & -44.9 & 0.25 & 1 \\
641 & 12:26:59.28 & 33:32:40.94 & 2 & 22.08 & -0.291 & 22.62 & 22.11 & -0.314 & 22.53 & 3.82 & 11.4 & 0.34 & 1 \\
647 & 12:27:00.00 & 33:32:40.87 & 1 & 22.27 & -0.434 & 22.09 & 22.25 & -0.417 & 22.16 & 4.16 & 23.6 & 0.19 & 1 \\
648 & 12:26:50.41 & 33:32:41.65 & 1 & 21.69 & -0.502 & 21.16 & 21.60 & -0.430 & 21.44 & 4.68 & -33.5 & 0.30 & 1 \\
649 & 12:26:56.33 & 33:32:41.24 & 2 & 22.48 & -0.607 & 21.44 & 22.38 & -0.525 & 21.75 & 4.89 & 52.4 & 0.28 & 0 \\
650 & 12:27:10.57 & 33:32:41.90 & 1 & 22.24 & -0.417 & 22.14 & 22.30 & -0.466 & 21.96 & 3.56 & -35.6 & 0.19 & 1 \\
656 & 12:26:56.72 & 33:32:40.68 & 2 & 22.02 & -0.758 & 20.22 & 21.96 & -0.707 & 20.41 & 4.75 & -15.8 & 0.09 & 0 \\
675 & 12:26:54.68 & 33:32:35.61 & 2 & 22.34 & -0.685 & 20.90 & 22.38 & -0.709 & 20.82 & 4.30 & 67.8 & 0.50 & 1 \\
685 & 12:26:54.80 & 33:32:34.43 & 2 & 22.63 & -0.524 & 22.00 & 22.68 & -0.561 & 21.87 & 3.66 & -8.4 & 0.34 & 0 \\
689 & 12:26:55.30 & 33:32:32.78 & 2 & 22.29 & -0.761 & 20.48 & 22.24 & -0.716 & 20.65 & 4.60 & 7.4 & 0.40 & 1 \\
703 & 12:26:56.12 & 33:32:23.37 & 1 & 20.60 & -0.094 & 22.13 & 20.65 & -0.099 & 22.15 & 4.54 & -42.1 & 0.11 & 1 \\
709 & 12:26:57.13 & 33:32:27.83 & 1 & 21.90 & -0.857 & 19.60 & 21.91 & -0.869 & 19.55 & 4.32 & -70.9 & 0.55 & 1 \\
711 & 12:27:00.52 & 33:32:30.14 & 1 & 23.45 & -0.906 & 20.91 & 23.71 & -1.055 & 20.43 & 1.18 & 31.1 & 0.70 & 1 \\
739 & 12:27:04.09 & 33:30:56.10 & 1 & 21.78 & -0.194 & 22.80 & 21.94 & -0.318 & 22.34 & 3.14 & 27.2 & 0.32 & 0 \\
757\tablenotemark{\dag} & 12:27:04.65 & 33:31:01.17 & 1 & 22.49 & -0.529 & 21.83 & 22.87 & -0.785 & 21.02 & 1.65 & -80.4 & 0.67 & 1 \\
760 & 12:27:03.83 & 33:31:01.45 & 1 & 21.78 & -0.738 & 20.08 & 21.74 & -0.708 & 20.19 & 4.47 & -49.8 & 0.47 & 1 \\
781 & 12:26:46.73 & 33:31:03.83 & 1 & 22.07 & -0.888 & 19.61 & 22.05 & -0.886 & 19.61 & 4.85 & -37.0 & 0.33 & 0 \\
798 & 12:26:56.10 & 33:31:09.63 & 1 & 24.37 & -1.142 & 20.65 & 24.71 & -1.240 & 20.50 & 0.26 & -31.8 & 0.56 & 0 \\
801 & 12:27:07.23 & 33:31:15.87 & 1 & 21.54 & -0.242 & 22.32 & 21.53 & -0.235 & 22.35 & 4.07 & 87.2 & 0.35 & 1 \\
805\tablenotemark{\dag} & 12:27:04.35 & 33:31:21.61 & 1 & 21.21 & -0.069 & 22.86 & 21.81 & -0.492 & 21.34 & 1.06 & 47.2 & 0.65 & 0 \\
841\tablenotemark{\dag} & 12:27:04.49 & 33:31:32.86 & 1 & 20.78 & 0.025 & 22.90 & 21.30 & -0.365 & 21.47 & 0.85 & 30.9 & 0.73 & 0 \\
861 & 12:27:04.90 & 33:31:32.46 & 1 & 20.93 & -0.392 & 20.96 & 21.20 & -0.575 & 20.31 & 1.89 & -49.7 & 0.62 & 0 \\
863\tablenotemark{\dag} & 12:27:00.87 & 33:31:27.15 & 2 & 21.28 & -0.181 & 22.37 & 21.48 & -0.329 & 21.82 & 2.75 & 0.8 & 0.63 & 0 \\
872\tablenotemark{\dag} & 12:26:58.08 & 33:31:34.58 & 1 & & & & 22.38 & -0.128 & 23.73 & 1.46 & -0.8 & 0.87 & 0 \\   
883 & 12:26:56.32 & 33:31:37.20 & 1 & 22.76 & -0.669 & 21.41 & 22.91 & -0.777 & 21.02 & 2.70 & -48.4 & 0.58 & 1 \\
899 & 12:26:45.63 & 33:31:40.66 & 1 & 22.05 & -0.780 & 20.14 & 22.06 & -0.785 & 20.12 & 3.90 & -25.9 & 0.47 & 1 \\
907\tablenotemark{\dag} & 12:27:05.19 & 33:31:42.80 & 1 & 21.09 & 0.382 & 24.99 & 21.91 & -0.170 & 23.05 & 0.63 & -16.3 & 0.77 & 0 \\
910 & 12:26:45.62 & 33:31:41.79 & 1 & 22.49 & -0.664 & 21.16 & 22.45 & -0.629 & 21.29 & 4.42 & 81.9 & 0.40 & \\ 
928\tablenotemark{\dag} & 12:27:09.15 & 33:31:47.10 & 1 & 19.62 & -0.066 & 21.28 & 19.87 & -0.257 & 20.57 & 1.95 & -29.5 & 0.70 & 0 \\
933 & 12:26:52.98 & 33:31:46.86 & 1 & 22.76 & -0.925 & 20.12 & 22.81 & -0.955 & 20.02 & 3.37 & 17.0 & 0.61 & 0 \\
934\tablenotemark{\dag} & 12:26:53.53 & 33:31:50.23 & 1 & 19.83 & 0.500 & 24.32 & 20.78 & -0.069 & 22.43 & 1.02 & 43.8 & 0.24 & 0 \\
960\tablenotemark{\dag} & 12:27:04.98 & 33:31:56.17 & 1 & 20.45 & 0.402 & 24.46 & 21.39 & -0.225 & 22.25 & 0.65 & 31.2 & 0.50 & 0 \\
968 & 12:26:55.83 & 33:31:53.45 & 1 & 22.27 & -0.588 & 21.32 & 22.33 & -0.640 & 21.12 & 3.46 & -45.2 & 0.56 & 0 \\
982\tablenotemark{\dag} & 12:27:03.02 & 33:31:57.27 & 1 & 18.92 & 0.160 & 21.71 & 19.30 & -0.119 & 20.69 & 2.17 & -60.2 & 0.35 & 0 \\
995 & 12:26:46.13 & 33:32:01.54 & 2 & 21.45 & -0.406 & 21.41 & 21.44 & -0.397 & 21.44 & 4.09 & 28.0 & 0.24 & 0 \\
996\tablenotemark{\dag} & 12:26:55.64 & 33:32:13.09 & 1 & 20.10 & -0.178 & 21.20 & 20.47 & -0.438 & 20.27 & 1.28 & -25.4 & 0.84 & 1 \\
999 & 12:26:51.27 & 33:32:05.18 & 1 & 22.41 & -0.377 & 22.52 & 22.72 & -0.603 & 21.70 & 2.12 & 85.4 & 0.30 & 1 \\
1001\tablenotemark{\dag} & 12:26:59.24 & 33:32:12.62 & 2 & 20.50 & 0.551 & 25.24 & 21.34 & 0.007 & 23.37 & 1.50 & 76.6 & 0.27 & 0 \\
1005\tablenotemark{\dag} & 12:26:54.82 & 33:32:21.81 & 1 & 22.24 & -0.235 & 23.05 & 22.10 & -0.369 & 22.25 & 0.54 & -83.4 & 0.67 & 1 \\
1009\tablenotemark{\dag} & 12:26:55.55 & 33:32:17.60 & 1 & 21.45 & 0.576 & 26.32 & 22.42 & -0.027 & 24.27 & 0.88 & -46.9 & 0.66 & \\ 
1022\tablenotemark{\dag} & 12:27:08.20 & 33:32:07.47 & 1 & 24.89 & -0.040 & 26.68 & 23.65 & -0.347 & 23.91 & 0.44 & -5.2 & 0.22 & 1 \\
1025 & 12:26:56.57 & 33:32:19.72 & 1 & 22.56 & -0.702 & 21.04 & 22.26 & -0.443 & 22.04 & 6.96 & -92.4 & 0.50 & 1 \\
1027\tablenotemark{\dag} & 12:27:03.68 & 33:32:10.55 & 1 & 22.20 & -0.088 & 23.76 & 22.89 & -0.554 & 22.11 & 0.64 & 51.5 & 0.74 & 0 \\
1042\tablenotemark{\dag} & 12:26:53.54 & 33:32:19.27 & 1 & 22.03 & 0.085 & 24.44 & 22.80 & -0.436 & 22.61 & 0.82 & 41.8 & 0.64 & 0 \\
1047 & 12:26:56.08 & 33:32:15.75 & 1 & 21.76 & -0.764 & 19.92 & 21.81 & -0.802 & 19.79 & 3.30 & -4.8 & 0.60 & 1 \\
1057\tablenotemark{\dag} & 12:27:05.52 & 33:30:43.90 & 1 & 22.41 & 0.022 & 24.52 & 23.13 & -0.457 & 22.83 & 1.02 & -31.5 & 0.16 & 0 \\
1080\tablenotemark{\dag} & 12:27:01.74 & 33:30:25.27 & 2 & 21.90 & -0.214 & 22.82 & 22.38 & -0.549 & 21.62 & 1.30 & -26.6 & 0.72 & 0 \\
1083\tablenotemark{\dag} & 12:26:49.64 & 33:30:19.40 & 1 & 21.84 & -0.159 & 23.04 & 21.62 & -0.089 & 23.17 & 3.26 & 87.3 & 0.41 & 0 \\
1091\tablenotemark{\dag} & 12:27:08.23 & 33:32:13.13 & 1 & 21.50 & 0.148 & 24.23 & 22.05 & -0.235 & 22.87 & 1.67 & -86.9 & 0.40 & 1 \\
1103\tablenotemark{\dag} & 12:27:02.61 & 33:30:13.75 & 2 & 22.15 & -0.022 & 24.03 & 22.84 & -0.496 & 22.36 & 0.72 & -0.6 & 0.73 & 0 \\
1157\tablenotemark{\dag} & 12:27:03.56 & 33:30:14.19 & 1 & 22.13 & 0.105 & 24.65 & 22.93 & -0.434 & 22.75 & 0.69 & 40.5 & 0.65 & 0 \\
1164 & 12:26:55.69 & 33:30:37.88 & 1 & 22.18 & -0.442 & 21.97 & 22.42 & -0.620 & 21.31 & 2.31 & 50.4 & 0.15 & 1 \\
1170 & 12:26:49.39 & 33:30:31.35 & 1 & 21.36 & -0.328 & 21.71 & 21.34 & -0.313 & 21.76 & 4.12 & 24.7 & 0.25 & 1 \\
1175 & 12:26:57.81 & 33:30:37.75 & 1 & 21.68 & -0.655 & 20.39 & 21.72 & -0.682 & 20.30 & 4.44 & 78.6 & 0.25 & 0 \\
1196\tablenotemark{\dag} & 12:27:01.79 & 33:30:22.25 & 2 & 22.37 & 0.059 & 24.66 & 23.16 & -0.473 & 22.79 & 0.66 & 53.9 & 0.72 & 0 \\
1199 & 12:27:08.57 & 33:30:48.16 & 1 & 22.13 & -0.748 & 20.38 & 22.07 & -0.699 & 20.57 & 4.71 & 79.8 & 0.22 & 1 \\
1251 & 12:26:56.33 & 33:32:41.81 & 2 & 22.38 & -0.655 & 21.10 & 22.60 & -0.797 & 20.60 & 1.90 & 65.3 & 0.74 & 1 \\
1252 & 12:27:03.15 & 33:31:55.36 & 1 & 21.89 & -0.218 & 22.79 & 22.16 & -0.407 & 22.11 & 2.74 & -2.9 & 0.34 & 1 \\
1253\tablenotemark{\dag} & 12:27:07.70 & 33:32:54.26 & 1 & 21.77 & 0.234 & 24.93 & 22.80 & -0.389 & 22.85 & 0.46 & -3.8 & 0.72 & \\ 
1254 & 12:26:56.12 & 33:31:09.35 & 1 & 21.97 & -0.065 & 23.64 & 21.90 & -0.014 & 23.82 & 4.31 & -53.9 & 0.30 & \\

\enddata
                                                                                                                                                         
\tablenotetext{a}{1: galaxy is considered a member of RX J1226.9+3332; 0: galaxy is not a member,
left blank indicates redshift not available}
\tablenotetext{\dag}{Galaxy exhibits late-type structure and is not well fit with
an r$^{1/4}$ law profile.}
\end{deluxetable}

\clearpage

\clearpage
\begin{deluxetable}{llllrrrrrrrrrl}
\rotate
\tabletypesize{\scriptsize}
\tablewidth{0pt}
\tablecaption{RXJ0142.0+2131: Photometric and Structural Parameters\label{rxj0142cat}}
\tablehead{
\colhead{ID} &
\colhead{RA} &
\colhead{Dec} &
\colhead{N$_{meas}$} &
\colhead{m$_{tot,dev}$} &
\colhead{$\log(r_{e})_{dev}$} &
\colhead{$\langle\mu \rangle_{e,dev}$} &
\colhead{m$_{tot,ser}$} &
\colhead{$\log(r_{e})_{ser}$} &
\colhead{$\langle\mu \rangle_{e,ser}$} &
\colhead{n$_{ser}$} &
\colhead{PA$_{ave}$} &
\colhead{$\epsilon_{ave}$} &
\colhead{Member\tablenotemark{a}} \\
\colhead{} & 
\colhead{(2000)} &
\colhead{} &
\colhead{} &
\colhead{(F775W)} &
\colhead{(arcsec)} &
\colhead{mag arcsec$^{-2}$} &
\colhead{(F775W)} &
\colhead{(arcsec)} &
\colhead{mag arcsec$^{-2}$} &
\colhead{} &
\colhead{} &
\colhead{} &
\colhead{} \\
}
\startdata

1 & 1:42:09.11 & 21:33:23.84 & 1 & 17.46 & 0.360 & 21.24 & 16.93 & 0.794 & 22.89 & 6.98 & -57.7 & 0.07 & 1 \\
22 & 1:42:08.68 & 21:33:22.62 & 1 & 18.98 & -0.313 & 19.40 & 18.97 & -0.305 & 19.43 & 4.09 & -85.9 & 0.28 & 1 \\
88 & 1:42:09.21 & 21:33:13.97 & 1 & 21.74 & -0.611 & 20.67 & 22.04 & -0.814 & 19.96 & 1.48 & 53.8 & 0.50 & 1 \\
128 & 1:42:07.37 & 21:33:02.13 & 1 & 18.76 & 0.029 & 20.90 & 18.80 & 0.002 & 20.80 & 3.77 & -0.2 & 0.21 & 1 \\
205 & 1:42:04.40 & 21:32:39.82 & 1 & 18.47 & 0.015 & 20.54 & 18.24 & 0.209 & 21.28 & 5.72 & 39.4 & 0.34 & 1 \\
318 & 1:41:57.29 & 21:32:27.84 & 1 & 19.73 & -0.058 & 21.43 & 20.04 & -0.294 & 20.56 & 1.91 & -14.7 & 0.41 & 1 \\
322 & 1:42:03.69 & 21:32:15.34 & 1 & 18.58 & 0.070 & 20.92 & 18.85 & -0.142 & 20.14 & 1.91 & -19.2 & 0.65 & 1 \\
379 & 1:42:01.92 & 21:32:10.19 & 1 & 18.71 & -0.276 & 19.32 & 18.68 & -0.252 & 19.41 & 4.30 & -37.4 & 0.52 & 1 \\
412 & 1:42:02.45 & 21:31:57.59 & 1 & 18.25 & -0.061 & 19.93 & 18.19 & -0.019 & 20.09 & 4.40 & -47.0 & 0.09 & 1 \\
442 & 1:42:02.64 & 21:32:06.54 & 1 & 20.74 & -0.539 & 20.04 & 20.87 & -0.634 & 19.69 & 2.77 & -33.0 & 0.32 & 1 \\
479 & 1:42:03.46 & 21:31:17.36 & 2 & 16.31 & 0.925 & 22.92 & 16.32 & 0.919 & 22.91 & 4.01 & -58.6 & 0.38 & 1 \\
537 & 1:42:08.64 & 21:31:45.50 & 1 & 19.97 & -0.461 & 19.65 & 19.86 & -0.370 & 20.00 & 5.07 & 6.6 & 0.21 & 1 \\
614 & 1:42:01.26 & 21:31:31.98 & 1 & 19.68 & -0.525 & 19.05 & 19.77 & -0.594 & 18.80 & 2.95 & 52.3 & 0.33 & 1 \\
637 & 1:42:01.38 & 21:31:22.13 & 1 & 19.41 & -0.246 & 20.17 & 19.43 & -0.266 & 20.09 & 3.78 & 19.5 & 0.32 & 1 \\
671 & 1:42:03.21 & 21:31:11.90 & 2 & 17.90 & -0.142 & 19.18 & 18.00 & -0.223 & 18.88 & 3.24 & 25.7 & 0.60 & 1 \\
760 & 1:42:01.28 & 21:31:04.59 & 2 & 19.87 & -0.566 & 19.03 & 19.93 & -0.608 & 18.88 & 3.22 & 3.7 & 0.48 & 1 \\
777 & 1:41:59.76 & 21:30:57.91 & 2 & 19.11 & -0.364 & 19.28 & 19.16 & -0.401 & 19.14 & 3.54 & -6.9 & 0.44 & 1 \\
844 & 1:42:07.17 & 21:30:49.71 & 1 & 19.69 & -0.437 & 19.49 & 19.65 & -0.409 & 19.60 & 4.37 & -76.7 & 0.47 & 1 \\
911 & 1:42:03.11 & 21:30:31.59 & 1 & 20.67 & -0.579 & 19.76 & 20.65 & -0.561 & 19.83 & 4.25 & 80.1 & 0.31 & 1 \\
1012 & 1:42:01.75 & 21:30:17.39 & 1 & 20.23 & -0.500 & 19.72 & 20.19 & -0.473 & 19.82 & 4.35 & 73.9 & 0.13 & 1 \\
1029 & 1:41:55.20 & 21:30:12.08 & 1 & 19.85 & -0.278 & 20.46 & 19.82 & -0.249 & 20.56 & 4.25 & 45.0 & 0.25 & 1 \\
1043 & 1:41:58.57 & 21:30:01.91 & 1 & 18.69 & -0.153 & 19.91 & 18.70 & -0.163 & 19.87 & 3.89 & 5.7 & 0.50 & 1 \\
1099 & 1:42:05.63 & 21:30:03.39 & 1 & 21.06 & -0.068 & 22.71 & 21.58 & -0.451 & 21.31 & 1.06 & 52.2 & 0.73 & 1 \\
1179 & 1:42:00.91 & 21:29:41.60 & 1 & 18.93 & 0.352 & 22.68 & 19.43 & -0.014 & 21.35 & 1.37 & -26.2 & 0.29 & 1 \\
1205 & 1:41:53.41 & 21:29:26.73 & 1 & 19.34 & -0.351 & 19.57 & 19.33 & -0.342 & 19.60 & 4.10 & -19.3 & 0.64 & 1 \\
1207 & 1:42:04.04 & 21:29:35.51 & 1 & 19.81 & -0.403 & 19.78 & 19.77 & -0.373 & 19.90 & 4.33 & 60.2 & 0.46 & 1 \\
1412 & 1:42:07.28 & 21:28:56.51 & 1 & 19.79 & -0.239 & 20.59 & 19.64 & -0.138 & 20.95 & 4.80 & 25.6 & 0.14 & 1 \\
1416 & 1:42:06.40 & 21:28:38.73 & 1 & 19.99 & 0.155 & 22.76 & 20.43 & -0.194 & 21.45 & 1.61 & -26.8 & 0.70 & 1 \\
\enddata
                                                                                                                                                         
\tablenotetext{a}{1: galaxy is considered a member of RX J0142.0+2131; 0: galaxy is not a member}
\end{deluxetable}

\clearpage

\clearpage
\begin{deluxetable}{lrrrr}
\tabletypesize{\normalsize}
\tablewidth{0pt}
\tablecaption{PSF Tests (Recovered - Input values)\label{psfresults}}
\tablehead{
\colhead{} &
\colhead{} &
\colhead{PSF} &
\colhead{} &
\colhead{} \\
\colhead{Parameter} &
\colhead{Real\tablenotemark{a}} &
\colhead{Raw TinyTim} &
\colhead{drizzled 3$^{\prime\prime}$} &
\colhead{drizzled 9$^{\prime\prime}$} \\
}
\startdata
$\langle\Delta(\log r_{e} + 0.8 \log \langle I\rangle_{e})\rangle_{r1/4}$ & $0.03\pm0.09$ & $0.05\pm0.10$ &
$0.03\pm0.10$ & $0.02\pm0.09$ \\
$\langle\Delta(\log r_{e} + 0.8 \log \langle I\rangle_{e})\rangle_{ser}$ & $0.01\pm0.05$ & $0.01\pm0.04$ &
$0.01\pm0.06$ & $-0.01\pm0.04$ \\
$\langle\Delta mag\rangle_{r1/4}$ & $0.2\pm0.3$ & $0.3\pm0.3$ & $0.2\pm0.3$ &
$0.2\pm0.3$ \\
$\langle\Delta mag\rangle_{ser}$ & $-0.1\pm0.2$ & $-0.1\pm0.1$ & $-0.1\pm0.2$ & $-0.0\pm0.2$ \\
$\langle\Delta \log r_{e}\rangle_{r1/4}$ & $-0.2\pm0.2$ & $-0.2\pm0.2$ & $-0.1\pm0.2$ & $-0.1\pm0.2$ \\
 $\langle\Delta \log r_{e}\rangle_{ser}$ & $0.0\pm0.1$ & $0.1\pm0.1$ & $0.0\pm0.1$
& $0.0\pm0.1$ \\
\enddata

\tablecomments{Values listed are the average offset and standard deviation of
the measurements.}
\tablenotetext{a}{Empirical PSFs generated from real stars}
\end{deluxetable}

\clearpage

\clearpage
\begin{deluxetable}{lrrrr}
\tabletypesize{\normalsize}
\tablewidth{0pt}
\tablecaption{Profile Fitting Tests (Recovered - Input values)\label{devresults}}
\tablehead{
\colhead{Parameter} &
\colhead{z=0.83} &
\colhead{z=0.28} &
\colhead{z=0.83} &
\colhead{z=0.28}  \\
\colhead{} &
\colhead{All sims} &
\colhead{All sims} &
\colhead{n$_{ser} > 2$} &
\colhead{n$_{ser} > 2$}  \\
}
\startdata

$\bar{\Delta} mag_{r1/4}$ & $-0.19\pm0.41$ & $-0.38\pm0.43$ & $-0.08\pm0.27$ & $-0.12\pm0.28$ \\
$\bar{\Delta}\log r_{e \ r1/4}$ & $0.12\pm0.24$ & $0.25\pm0.34$ & $0.05\pm0.17$ & $0.09\pm0.21$ \\
$\bar{\Delta} \langle\mu\rangle_{e \ r1/4}$ & $0.43\pm0.85$ & $0.87\pm1.41$ & $0.18\pm0.62$ & $-0.32\pm1.02$ \\
$\bar{\Delta}$ FPP$_{r1/4}$ & $-0.01\pm0.06$ & $-0.04\pm0.10$ & $-0.01\pm0.05$ & $-0.02\pm0.10$ \\
   &    &    &   &  \\
$\bar{\Delta} mag_{ser}$ & $0.01\pm0.30$ & $-0.04\pm0.46$  & $0.02\pm0.32$ & $-0.04\pm0.52$ \\
$\bar{\Delta} \log r_{e \ ser}$ & $-0.01\pm0.19$ & $0.02\pm0.23$ & $-0.01\pm0.21$ & $0.02\pm0.23$  \\
$\bar{\Delta} \langle\mu\rangle_{e \ ser}$ & $-0.06\pm0.73$ & $0.03\pm1.10$  & $-0.07\pm0.81$ & $0.03\pm1.20$ \\
$\bar{\Delta}$ FPP$_{ser}$ & $0.00\pm0.05$ & $0.00\pm0.09$ & $0.01\pm0.05$ & $0.01\pm0.10$ \\
$\bar{\Delta}$ n$_{ser}$ & $-0.2\pm1.1$ & $0.2\pm1.7$ & $-0.2\pm1.3$ & $0.2\pm1.9$ \\
   &    &    &   &  \\
$\bar{\Delta} mag_{BDr1/4}$ & $-0.56\pm0.67$ &  &  &  \\
$\bar{\Delta} \log r_{e \ BDr1/4}$ & $0.19\pm0.24$ &  &  &  \\
$\bar{\Delta} \langle\mu\rangle_{e \ BDr1/4}$ & $0.37\pm0.61$ &  &  &  \\
$\bar{\Delta}$ FPP$_{BDr1/4}$ & $0.07\pm0.09$ &  &  &  \\

\enddata

\tablecomments {r1/4 refers to profile fits with r$^{1/4}$ law profiles, ser to Sersic profile
fits, and BDr1/4 to objects generated with 2 components but fit with a single r$^{1/4}$ law
profile.  Errors listed are the rms scatter in the differences.
} 

\end{deluxetable}

\clearpage

\clearpage
\begin{deluxetable}{lllll}
\tabletypesize{\normalsize}
\tablewidth{0pt}
\tablecaption{r$^{1/4}$ Law Profile Fitting Tests (Recovered - Input values)\label{fppresults}}
\tablehead{
\colhead{} &
\colhead{} &
\colhead{$\langle\Delta$FPP$\rangle$} &
\colhead{} &
\colhead{} \\
\colhead{} &
\colhead{range1} &
\colhead{range2} &
\colhead{range3} \\
}
\startdata
Nearest neighbor & d $< 2^{\prime\prime}$ & $2 < $ d $ < 6^{\prime\prime}$ & d $> 6^{\prime\prime}$ \\
 & $-0.01\pm0.12$ & $-0.00\pm0.03$ & $0.00\pm0.04$  \\

Size &  $\log r_e > 0$  & $-1 < \log r_e < 0$ & $\log r_e < -1$  \\
   & $-0.00\pm0.05$ & $-0.01\pm0.05$ &  $-0.01\pm0.10$  \\

Mag & $i^{\prime} < 21$ & $21 < i^{\prime} < 22.5$  & $i^{\prime} > 22.5$ \\
 & $0.00\pm0.04$ & $-0.00\pm0.03$ & $-0.01\pm0.07$ \\

$\langle\mu\rangle_e$ & $\langle\mu\rangle_e < 20$ &  $20 < \langle\mu\rangle_e < 22.5$   &  $\langle\mu\rangle_e > 22.5$ \\
   &  $-0.00\pm0.08$ & $-0.01\pm0.05$ & $-0.00\pm0.05$ \\

\enddata
\end{deluxetable}

\clearpage

\clearpage
\begin{deluxetable}{lrrrr}
\tabletypesize{\normalsize}
\tablewidth{0pt}
\tablecaption{Profile Fitting Tests (external comparison)\label{bkresults}}
\tablehead{
\colhead{Parameter} &
\colhead{(This work - Blakeslee et al.)\tablenotemark{a}} &
\colhead{} &
\colhead{(subset Blakeslee et al.} &
\colhead{$n=4$)\tablenotemark{b}} \\
\colhead{} &
\colhead{zero point shift applied} &
\colhead{no shift}  &
\colhead{This work: Sersic} &
\colhead{r$^{1/4}$ fits} \\
}

\startdata

$\bar{\Delta} mag_{ser}$ & $0.01\pm0.08$ & $-0.02\pm0.08$ & $-0.01\pm0.09$ & $0.01\pm0.05$ \\
$\bar{\Delta} \log r_{e \ ser}$ & $-0.01\pm0.06$ & $-0.01\pm0.06$ & $0.01\pm0.07$ & $-0.02\pm0.03$ \\
$\bar{\Delta} \langle\mu\rangle_{e \ ser}$ & $-0.02\pm0.21$ & $-0.04\pm0.21$ & $0.04\pm0.26$  & $-0.07\pm0.10$  \\
$\bar{\Delta} n$ & $0.2\pm0.4$ & $0.2\pm0.4$ & &   \\
$\bar{\Delta}$ FPP$_{ser}$ & $0.00\pm0.01$ & $0.01\pm0.01$ & $-0.003\pm0.014$ & $0.006\pm0.006$ \\

\enddata

\tablenotetext{a}{Comparison between 23 galaxies in common.}
\tablenotetext{b}{Comparison between 10 galaxies for which 
we find best Sersic fits with n $> 4.0$ while Blakeslee et al. limit n $\leq 4.0$.
We show the comparison with both our Sersic and r$^{1/4}$ law (n $= 4.0$) fits.}
\end{deluxetable}

\clearpage


\end{document}